\documentclass{aastex}          
\usepackage{spr-astr-addons}    
\def \grss {GRS 1915+105~}
\def \grs {GRS 1915+105}
\def \trec{$T_{rec}$}
\def \trecs{$T_{rec}$~}
\def \sax {{\it Beppo}SAX}
\def \Pul{{\it Pulse}}
\def \Puls{{\it Pulse~}}

\begin{document}

\title{Non-linear oscillator models for the X-ray bursting of the microquasar 
GRS~1915+105}

\shorttitle{Non-linear models for the X-ray bursting of GRS~1915+105}
\shortauthors{E. Massaro et al.}

\author{E.~Massaro\altaffilmark{1}}
\affil{Dipartimento di Fisica, Universit\`a La Sapienza, Piazzale A. Moro 2, 
I-00185 Roma, Italy}
\and \author{A.~Ardito\altaffilmark{2}} \and \author{P.~Ricciardi\altaffilmark{2}}
\affil{Dipartimento di Matematica, Universit\`a La Sapienza, Piazzale A. Moro 2, 
I-00185 Roma, Italy}
\and \author{F.~Massa\altaffilmark{3}}
\affil{INFN, Sezione Roma1, Universit\`a La Sapienza, Piazzale A. Moro 2, 
I-00185 Roma, Italy}
\and \author{T.~Mineo\altaffilmark{4}}
\affil{INAF, IASF Palermo, via U. La Malfa 153, I-90146 Palermo, Italy}
\and \author{A.~D'A\`i\altaffilmark{5}}
\affil{Dipartimento di Fisica, Universit\`a di Palermo, Via Archirafi 36, I-90123 Palermo, Italy}
\email{mineo@iasf-palermo.inaf.it} 

\begin{abstract}
The microquasar \grs, exhibits a large variety of characteristic states, 
according to its luminosity, spectral state, and variability. 
The most interesting one is the so-called $\rho$-state, whose light curve 
shows recurrent bursts. This paper  presents a model based on 
Fitzhugh-Nagumo equations containing two variables: 
$x$, linked to the source photon luminosity $L$ detected by the MECS, 
and $y$ related to the mean photon  energy.  
We aim at providing a simple mathematical framework composed by non-linear
differential equations useful to predict  the observed light curve and 
the energy lags for the $\rho$-state and possibly other classes of the source.
We studied the equilibrium state and the stability conditions of this system
that includes one external parameter, $J$, 
that can be considered a function of the disk accretion rate.
Our work is based on  observations performed with the MECS
on  board  \sax~  when the source was in $\rho$  and $\nu$ mode, respectively.
The  evolution of the mean count rate and  photon energy  
were derived from a study of the trajectories in  the count rate - photon energy plane.
Assuming  $J$  constant, we found a solution  that reproduces  the  $x$ profile of 
the $\rho$ class bursts and, unexpectedly, we found that 
$y$ exhibited a time modulation  similar to that of the mean energy.
Moreover, assuming a slowly modulated $J$  the solutions for 
$x$ quite similar to those observed in the $\nu$ class light curves  is reproduced.
According these results,  the outer mass accretion rate  is probably responsible
for the state transitions, but within the $\rho$-class it is constant. 
This finding makes stronger the heuristic meaning of the non-linear model and 
suggests a simple relation between the variable $x$ and $y$. 
However, how a system of dynamical equations can be derived from the complex
mathematical apparatus of accretion disks remains to be furtherly explored.
\end{abstract}

\keywords{binaries - stars: individual: GRS 1915+105 -
 X-rays: stars - black hole physics - dynamical system}

\section{Introduction}
 
It is known that phenomena occurring in accretion disks around black holes
involve non-linear processes whose evolution can be described by a system 
of differential equations containing several quantities not directly observable.
The only information we have concerns a fraction of the dissipated energy 
via electromagnetic radiation and usually observed in a rather limited frequency 
band.
The stability of disk structures is also a very interesting subject 
of investigations since many years and theoretical analysis suggested that 
thermal and viscous instabilities can develop and establish a limit cycle 
behaviour.

To now the most important X-ray source exhibiting a complex variability, 
that on some occasions were characterized by long series of bursts as those 
expected by a limit cycle is the bright microquasar \grs, discovered by 
\citet{Castrotirado1992}.
Only recently, \citet{Altamirano2011} reported the discovery of IGR J17091+3624 
that exhibits variability patterns very similar to those of \grs.

The large variety of light curves of \grs, changing from quiescent states to
fast series of short bursts and to much more complex patterns of alternating 
bursting and quiescent phases was classified in 12 types by \citet{Belloni2000} 
on the basis of a large collection of multi-epoch RXTE observations.
New classes were added to these in the following years 
\citep{Naik2002,Hannikainen2003,Hannikainen2005}
indicating that the source is potentially able to develop a rather large number 
of physical conditions from which more types of light curves can be originated.
A description of such a complex phenomenology is given in the review paper by
\citet{Fender2004}.

From a general point of view the light curve variability classes of \citet{Belloni2000}
can be grouped in three main types: 
$i$) light curves characterised only by small amplitude noisy fluctuations 
with respect to a stable average level (e.g. classes $\phi$, $\chi$, and $\delta$);
$ii$) light curves presenting series of (positive or negative) pulses 
(e.g. classes $\gamma$, $\kappa$, and $\rho$), occasionally exhibiting a rather stable 
recurrence time; $iii$) light curves structured in sequences of fast spikes alternating 
with rather quiescent and low brightness states (e.g. classes $\theta$, $\lambda$, 
$\alpha$, $\beta$ and $\nu$).

One of the most interesting variability classes is the $\rho$,
lasting several days, whose light curves 
are quasi-regular series of bursts with a moderately variable recurrence time, 
usually in the range 40 -- 100 seconds.
The time and spectral properties of $\rho$ class bursts have been investigated by
several authors and the most recent papers on this subject are those by 
\citet{Neilsen2011, Neilsen2012} 
based on RXTE data and those by \citet{Massaro2010}, \citet{Mineo2012}, and 
\citet{Massa2013} who considered a long observation performed with \sax~ in October 2000.
In particular,  \citet{Massaro2010} reported that the mean recurrence time of the 
bursts increases with the source brightness, while \citet{Massa2013} investigated 
the properties of loops described by $\rho$ bursts in a dynamical space where the 
coordinates are the count rate and the mean energy of photons. 
Since the first analysis  \citep{Taam1997} these recurrent bursts were 
associated with a limit cycle due to the onset of some disk instability. 
Several authors calculated possible theoretical light curves of the bolometric luminosity 
originating from disk instabilities.
Complexity of hydrodynamic and thermodynamic equations does not allow a rather simple 
picture of the roles played by the involved physical quantities and the interpretation
of data is not straightforward.
Moreover, the limit cycle is often described in terms of disk quantities, such as the
integrated density or the mass accretion rate, which are not directly observable

In the present paper we adopt a different approach and study the solutions of
non-linear systems of two and three ordinary differential equations, whose solutions  
have very close similarities with the observational data series.
These systems are mainly applied in the simulation of neuronal behaviour, and are
able to describe quiescent, spiking and bursting activity like the one exhibited by 
\grs.
We will show that this approach makes possible to calculate light curves and phase 
space trajectories useful to investigate some dynamical aspects of the instability
processes.

\begin{figure}[tb]
\includegraphics[width=1.0\columnwidth]{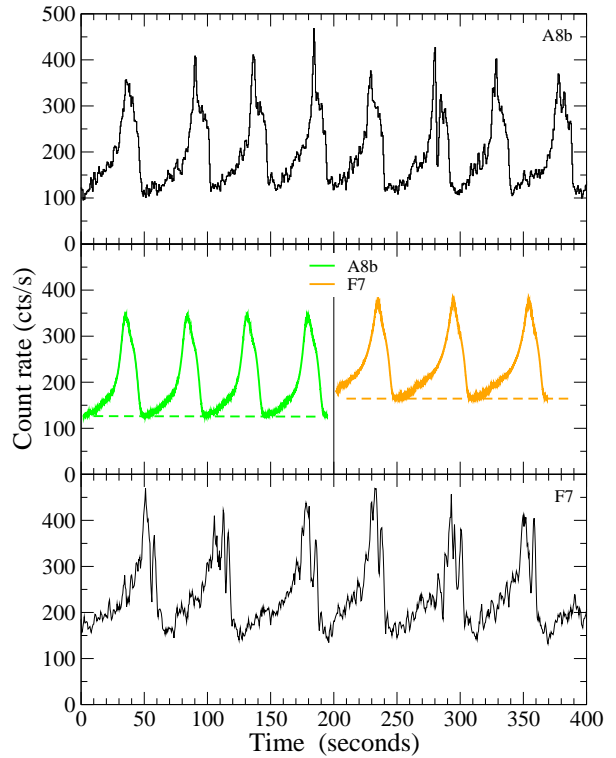}
\caption{
Two 300 second long segments of the count rate of the MECS [1.4 $-$ 10] keV 
for the data series A8b (top panel), F7 (bottom panel) of the long observation of October
2000 \citep[see][]{Massaro2010}. 
A running average smoothing over 3 bins is applied to reduce the statistical
Poissonian noise. 
The bin size of both series is 0.5 s.
In the central panel two curves reproducing the mean bursts' profile are shown:
the green one on the left corresponds to the A8b data and the orange curve on the right
to F7}
\label{fig1}
\end{figure}

Non-linear oscillators were already considered in stellar physics for describing the 
convective energy transfer \citep{Moore1966} and the dynamics of pulsating 
variable stars \citep{Regev1981, Buchler1981, Buchler1993}. 
Non-linear processes are also present in the coupling of the hot plasma and the radiation 
field in an accreting disk around a compact object.
It is, therefore, likely that such equations can represent a useful mathematical 
approximation of much more complex relations in a suitable neighborhood of an
equilibrium point.
In this paper, we will not deal with the physical description of an accretion disk but
will limit our study to show how the behaviour of \grss can be described by a unique 
oscillator and that some observed changes can be related to variations of a single
parameter.

In Sect. 2 we describe the coarse structure of X-ray bursts and our method to
compute the mean count rate and photon energy time curves.
In Sect. 3 non-linear oscillator models are introduced and a solution with only
two variables and constant parameters for the \grss data is presented; its 
equilibrium point and stability is studied in Sect. 4.
In Sect. 5 we investigate the consequences of parameters' changes, and in Sect. 6
a possible extension to dynamical systems with three equations is presented.
Finally, in Sect. 7 we discuss our results in the framework of current models for
disk instabilities and limit cycles.

\section{Time structure of $\rho$ class bursts}

We aim to demonstrate that the solutions of the equations of a rather simple non-linear 
oscillator are able to reproduce some of the main properties of the complex bursting patterns 
observed in \grs.
Considering that the most relevant information is mainly derived from the observed light
curves it is useful to describe the structure of the bursts and to define typical time 
scale ratios.
However, because of the variability of individual bursts, it is useful to use mean
burst profiles for defining the main sections considered in our analysis.
Mean pulse profiles and photon energy curves were obtained from a study of the 
trajectories in the count rate - mean photon energy ({\it CR-E}) plane. 
Fig.~\ref{fig1} shows two short segments of $\rho$ class light curves (named A8b and F7, 
see \citet{Massaro2010}, for the nomenclature and a description of the time 
properties of these data series) observed by the two operating MECS detectors on 
board the \sax~ satellite in October 2000.
These curves are representative of the first and third of the three time intervals in 
which \citet{Massaro2010} divided the whole observation: in the first interval \grss 
exhibited a regular behaviour with only small changes of the burst recurrence time 
\trec, the second interval showed an irregular behaviour, characterized by a peak 
multiplicity $\geq$ 3, whereas in the third interval the source exhibited again a 
nearly regular variations with the mean count rate and the \trec~ higher than in 
the first interval.

\begin{figure}[tb]
\includegraphics[height=8.cm,angle=-90]{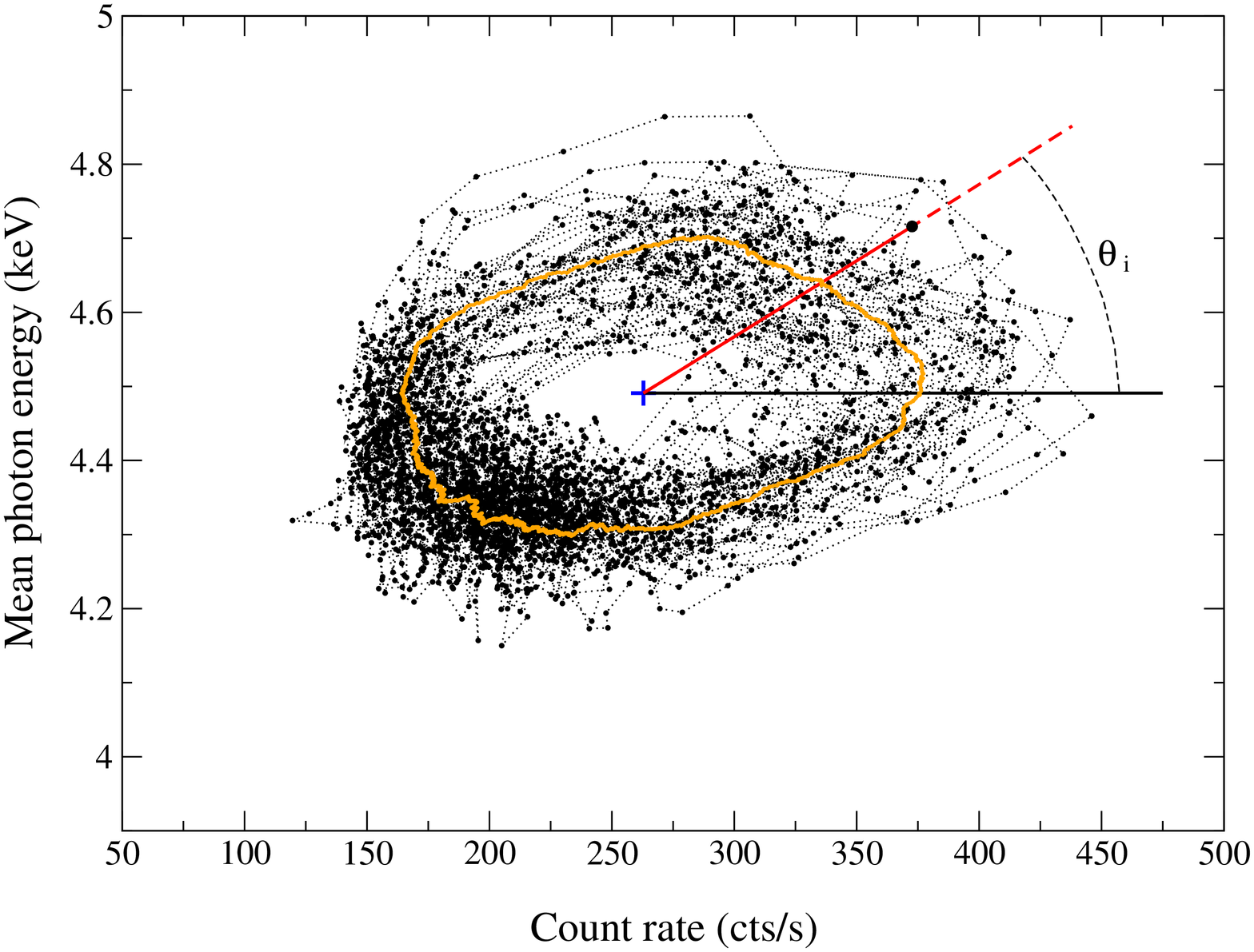}
\includegraphics[height=8.cm,angle=-90]{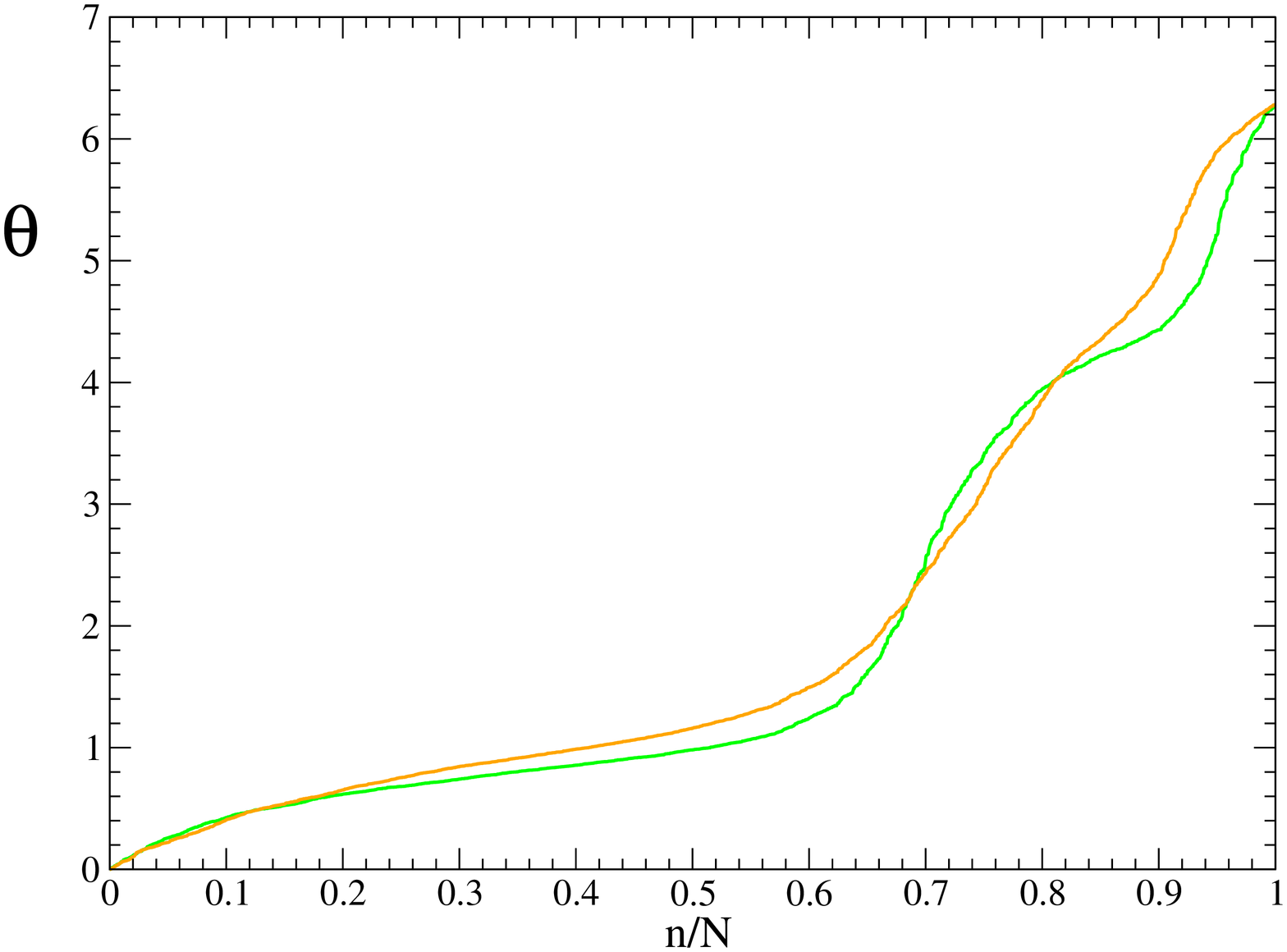}
\caption{
Upper panel:
Trajectory described by bursts of F7 series in the {\it CR-E} space.
The blue cross marks the centroid position. 
The red line is the segment connecting the centroid to a generic observed datum
$i$ (large dot) and the black line gives a direction from which the angle 
$\theta_i$ is measured.
The orange line is the resulting mean loop.
Lower panel: the plots of $\theta_n$ vs the ordering number $n$ (normalized to the
total number of data points in the series) for the A8b (green) and F7 (orange).
}
\label{fig2}
\end{figure}

\subsection{Mean burst profile reconstruction}

In our method, the starting point for obtaining the mean CR and photon energy series is
the analysis of loop trajectories described in the ({\it CR-E}) plane in the course of 
$\rho$ class bursts.
These trajectories were already described by \citet{Massa2013} and we 
we refer to that paper for the details of the applied algorithms. 
The upper panel in Fig.~\ref{fig2} shows the loops of the F7 series and the central cross marks 
the centroid position computed through an iterative process.

\begin{figure}[tb]
\hspace{-2cm}
\includegraphics[width=1.0\columnwidth]{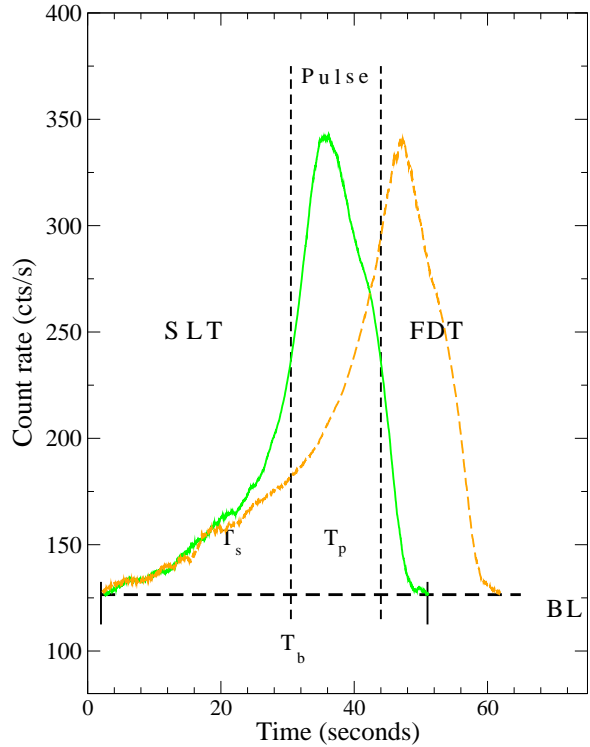}
\caption{
Coarse time structure of bursts typical of the $\rho$.
The green curve is the average profile of the A8b data series computed as explained in 
the text.
The three main segments, {\it SLT}, \Pul, and {\it FDT} refer to the green curve. 
The total burst time is $T_b$, while the durations of {\it SLT} and \Puls are $T_s$
and $T_p$, respectively.
{\it BL} indicates the stable base-line level over which bursts are superposed.
The dashed orange line is the mean pulse profile of the F7 data series translated to the
same BL level of the other data for a better comparison of the differences in their profiles.
}
\label{fig3}
\end{figure}

Once known the centroid coordinates and the \trec, obtained, for instance, by means of Fourier 
analysis, one considers a segment connecting the centroid with the $i$-th point of the data
series; it is so possible to define a phase angle $0 < \theta_i < 2 \pi $ with respect 
to a suitably chosen origin.
Data are then ordered according to the values of $\theta$ and a new ordering index $n$ is
obtained for the series and a running average is applied to smooth fluctuations until 
a regular profile is obtained.
The resulting plots of the ordered series $\theta_n$ vs $n$ for the two considered
series are given in the lower panel of Fig.~\ref{fig2}.
When time sampling index $i$ increases the segment rotates in the plane with a variable 
angular velocity that can be obtained from differences of the phase angles between their 
consecutive values. 
The length of considered data series ensures that they contain many bursts and therefore
the number of data points within any considered phase interval increases when the angular 
velocity is decreasing and viceversa.
It is reasonable to assume a proportionality between the time necessary to cover the phase 
interval $\Delta t / T_{rec}$ and the fraction of data $\Delta n/N$ falling inside.
One can thus compute the angular velocity of the vector as 
$(N/T_{rec}) (\Delta \theta / \Delta n)$. 
Finally, one can simply transform the time scale to write two data sets where count 
rates and mean photon energies are functions of the resulting fraction of \trec.
The mean pulse profiles of series A8b and F7 are shown in the central panels of Fig.~\ref{fig1}
and in Fig.~\ref{fig3}, the corresponding mean loop in the {\it CR-E} space is plotted in Fig.~\ref{fig2}.

\subsection{Burst structure}

The observed structure of the $\rho$ class bursts is rather complex and it is convenient
to distinguish between a {\it coarse} and a {\it fine} time structure, the latter clearly 
apparent when a time binning lower than about 1 second is used.
In particular the \Puls can contain two or more peaks of short duration, and in many cases
the first is significantly higher than the others. 
This fine structure, however, appears noisy and, because of the high count rates 
in the \sax~ data, it can be partially affected by some instrumental effects as telemetry 
limitations.
A good description of this fast and irregular phenomenon is quite difficult, whereas the
mean profile enveloping these narrow peaks is rather stable.  
Therefore, in the following of the present work we will focus our study on the large 
scale structure. 

According to \citet{Mineo2012}, mean profiles, like those shown in Fig.~\ref{fig3}, are divided 
into three main segments: 
the first is {\it Slow Leading Trail (SLT)} from the minimum level to half height,
the second one is the \Puls whose typical duration is measured by its HWHM, followed by 
the {\it Fast Decaying Tail (FDT)}, in which the count rate decreases to its minimum. 
$T_s$ that measures the duration of {\it SLT}, and $T_p$ that corresponds to the \Puls
length at the half height level, while the {\it FDT} duration is generally shorter than 
both these two covering a burst fraction of about 0.15 or less.
$T_b$ measures the total duration of individual bursts measured between two consecutive 
minima in the light curve.
We can thus define the two ratios:

\begin{equation}
r_s = T_s~/ T_b ~~~;~~~~~ r_p = T_p~/ T_b  
\end{equation}

\noindent
characterizing the durations of {\it SLT} and \Pul, respectively.
For example, for the mean A8b burst in Fig.~\ref{fig3} these ratios are $r_s = 0.58$ and $r_p = 0.28$,
but we recall that they can vary up to several percents in individual bursts.
In the case of the F7 mean burst, also plotted (orange curve) in Fig.~\ref{fig3} to show that the 
change in \trec~ is due to an increase of the {\it SLT} duration, we have $r_s = 0.63$ and 
$r_p = 0.25$, while mean duration and height of the \Puls were practically unchanged.

\begin{figure}[tb]
\includegraphics[height=8.2cm,angle=-90]{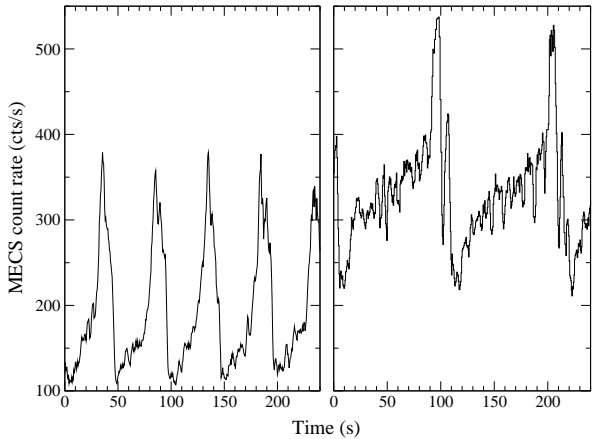}
\caption[]{
Two short segments of MECS light curves illustrating changes of bursts' structure.
Left panel: typical  $\rho$ class bursts from the A8b \sax~ data series of October 2000;
right panel: bursts of a data series in the final part of the same observation when the
variability class changed to $\nu$. 
Note the large difference in the BL count rate and of the recurring time. 
}
\label{fig4}
\end{figure}

In Fig.~\ref{fig1} it is apparent an increase of the count rate occurred on time 
scales much longer than the typical \trec: the lowest level is, in fact close to $\sim$120
cts/s, as for the A8b series,  while it increased to $\sim$160 cts/s in F7.
Moreover, the \trec~ of bursts increased by about 30\% \citep[see][for details]{Massaro2010}.

In addition to typical $\rho$ class bursts other types are observed.
For instance, relatively short sequences of bursts are present in other variability classes
like $\alpha$, $\beta$ and $\nu$ \citep{Belloni2000}, but their structure can present  
different features as shown in the two panel of Fig.~\ref{fig4}, where $\rho$ and $\nu$ bursts are
compared.
The latter ones have a quite narrow and sharp spike instead of a structured \Pul~and the 
leading trail is preceded by a fast initial increase just after a deep minimum.
Note also that when $\nu$ bursts were observed the lowest count rate level was higher
than that of the $\rho$ class bursts by a factor of about 2.

When considering burst profiles in different energy ranges as [1.3−3.4] and [6.8–10.2] keV
(see Fig. 2 in Massa et al. 2013) we found evidence of a few second delay of the higher energy
count rate with respect to the lower energy one, and this delay increases using PDS data, 
above 15 keV.
This Hard X-ray Delay (HXD) can be described by a change of the disk black body temperature
during the burst from $\sim$ 1 to $\sim$ 2 keV in the SLT and \Puls, respectively (Fig. 11 in
Massa et al. 2013).
This temperature change corresponds to a variation of the mean energy of the photons observed 
by MECS, that although limited in a rather narrow range, is able to depict the HXD in the 
{\it CR-E} plane.
The resulting loop structure in this plane was used by Massa et al. (2013) to estimate the
HXD.
Furthermore, the results of the spectral analysis presented by Mineo et al. (2012) indicated 
that the HXD cannot be interpreted as the effect of photon scattering in a hot corona.

\section{Dynamical equations of non-linear oscillators}

The complex behaviour of \grss presents many similarities with that of a
non-linear oscillating system as those used for describing signals in neuronal array.
Mathematical aspects of this important subject were deeply investigated in the
past half century and an extremely wide and technical literature is available.
We apply these methods to the study of the X-ray signals observed from \grss and
show that a rather simple, although non-linear, set of differential equations is
able to reproduce the properties of some variability classes by adjusting 
only one parameter. This can pave the way to define a frame based on the same
process useful to address theoretical models of disk oscillations and 
instabilities.

It is known from the theory of dynamical systems exhibiting either a quiescent 
or a spiking behaviour that they can be described by a system of first order 
non-linear differential equations.
In many case, two equations can provide a satisfactory picture of a rather large
class of phenomena, however, to take into account variation due to parameters
changing over different time scales, usually much longer than those typical of 
the system, a third (or a fourth) variable must be introduced. 
A simple harmonic oscillator with a damping (or growing) term and subject to a 
constant forcing is described by two variables whose first time derivatives are 
linear functions of them.

More generally, considering three dynamical variables $x$, $y$ and $z$, a rather 
general system having spiking and bursting solutions, thus involving changes on
different time scales, can be written as:

\begin{eqnarray}
\frac{dx}{dt} &=&  \frac{1}{A}[ P_3(x) - b_1 y - z] \nonumber \\
\frac{dy}{dt} &=& P_2(x) - b_2 y          \\      
\frac{dz}{dt} &=& \varepsilon [ s ( x -x_0) - z ]  \nonumber 
\end{eqnarray}

\noindent
where $P_2(x)$ and $P_3(x)$ are two polynomials of second and third degree,
respectively:

\begin{eqnarray}
P_3(x) &=& -a_1 x^3 + a_2 x^2 + a_3 x + a_4  \nonumber \\
P_2(x) &=& -a_5 x^2 + a_6 x + a_7 
\end{eqnarray}

For particular values of the parameters, the system of Eq.(2) can be 
reduced to well studied differential equation systems.
Thus for $a_2 = a_4 = a_5 = \varepsilon = 0$ we obtain the Fitzhugh-Nagumo 
(hereafter FhN) equations for only two variables, originally proposed by 
\citet{Fitzhugh1961}, who named it as Bonhoeffer-Van der Pol oscillator, and
extensively applied in simulating the behaviour of a neuron \citep[e.g.][]{Izhikevich2007}:

\begin{eqnarray}
\frac{dx}{dt} &=&  \frac{1}{A} [ -a_1 x^3 + a_3 x - b_1 y ] \nonumber \\
\frac{dy}{dt} &=& a_6 x - b_2 y + a_7  
\end{eqnarray}

\noindent
For the choice of parameters, $A= 1, a_3 = a_6 = 0, b_1 = -1, b_2 = 1$, 
Eq.(2) gives the Hindmarsh-Rose \citep[hereafter HR,][]{Hindmarsh1984} model 
 \citep[see also the tutorial paper by][]{Shilnikov2008}:

\begin{eqnarray}
\frac{dx}{dt} &=& -a_1 x^3 + a_2 x^2 + a_4 + y - z  \nonumber \\
\frac{dy}{dt} &=& -a_5 x^2 - y + a_7               \\
\frac{dz}{dt} &=& \varepsilon [ s ( x -x_0) - z ]  \nonumber 
\end{eqnarray}

An important characteristic of these equations is that non-linear terms imply 
that variables can evolve on different time scales. 
In the FhN model, for instance, $x$ is the {\it fast} variable and $y$ is 
{\it slow} one, whereas in the HR model $z$ is slow and the two others are fast.

\begin{figure}[tb]
\includegraphics[height=8.0cm,angle=-90,scale=1.0]{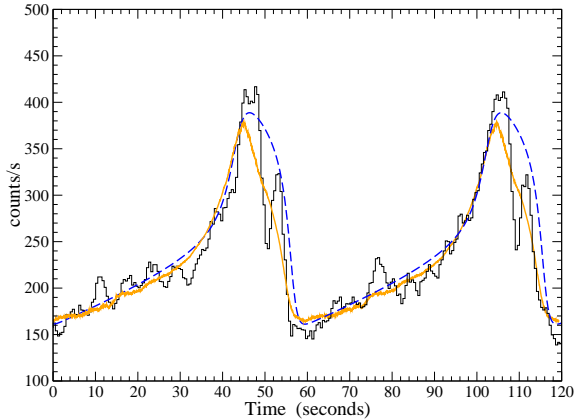}
\caption[]{
A short segment of the F7 MECS light curve showing two bursts (black) with 
a time binning of 0.5 s. 
Data are smoothed with a running average over 5 bins.
The orange curve is the mean reconstructed burst profile with the method
described in the text.
A solution of the FhN system, scaled and translated to match the data is
plotted as the blue long-dashed line. 
}
\label{fig5}
\end{figure}

\subsection{FhN model}
To reproduce the stable bursting pattern of regular $\rho$ class series, 
like the one observed in the A8b and F7 series (Fig.~\ref{fig1}), it is convenient to change
the variables to make simpler the FhN equations of Eq. (3) and to reduce the 
number of parameters.
These algebraic manipulations are described in the Appendix I; the resulting 
equations are: 

\begin{eqnarray}
\frac{dx}{dt} &=& - \rho x^3 + \chi x - \gamma y -  J \nonumber \\
\frac{dy}{dt} &=& x - y  
\end{eqnarray}

\noindent
where we indicated the two variables with $x$ and $y$, as before, and the
signs of the various terms were taken to have the parameters' values positive.

Parameters can be divided into {\it internal} and {\it external} ones if 
they appear  or not as factors of the variables.
We have thus only the external parameter $J$, whereas $\rho$, $\chi$ and $\gamma$
are internal.
Internal parameters are related to the physical state of the oscillating system,
while the external one can be considered as a forcing of the system.
In our first computations we will study the solutions obtained taking all the four 
parameters constant, although they, at least in principle, could change in time 
with the physical state of the source.
To take into account their possible changes one or more further equations must 
be added to the above system to make it autonomous.  
Effects of possible variations of the parameters and, in particular those originating
from slow changes of $J$ will be discussed in Sect. 5.

Note also that the particular form of Eqs. (6) does not offer a simple interpretation
for the physical meaning of the parameters.
This fact depends on the adopted form and to the fact that we are interested in
computing the time evolution of light curves.
As shown in the Appendix II, the system in Eq.(6) can be written in a more general
form, containing a quadratic term and an additional term to $J$, that leaves invariant the 
solutions for the two variables apart of a constant. 

Numerical computations were performed by means of a Runge-Kutta fourth order 
integration routine \citep{Press2007}.
Our first aim was to obtain a set of parameters' values for which the $x$ variable gave
a satisfactory solution for the F7 count rate light curve. 
Fig.~\ref{fig5} shows the resulting function compared with the mean burst profile (orange curve) 
and a short data segment.
The agreement, although not exact, is  fully satisfactory particularly for the {\it SLT}
shape: we obtained that $r_s =$ 0.64 and $r_p =$0.28 quite close to those given 
in the previous Section for the F7 series.
The adopted parameters' values were $\rho = 0.30$, $\chi = 33.0$, $\gamma = 222.0, J = 1100.0$. 
We remark here that these values were not obtained by means of a best fit optimization procedure, 
very hard to apply to the equation system solutions because of their high variability 
also for small changes of parameters.
Note that the model does not describe substructures in the \Pul, and that it is 
broader than observed close to the maximum; however, the {\it FDT} appears sharp
as in the data.
In the following we will refer to this solution as `FhN-A' model.

\begin{figure} [th!]
\includegraphics[height=7.5cm, angle=-90]{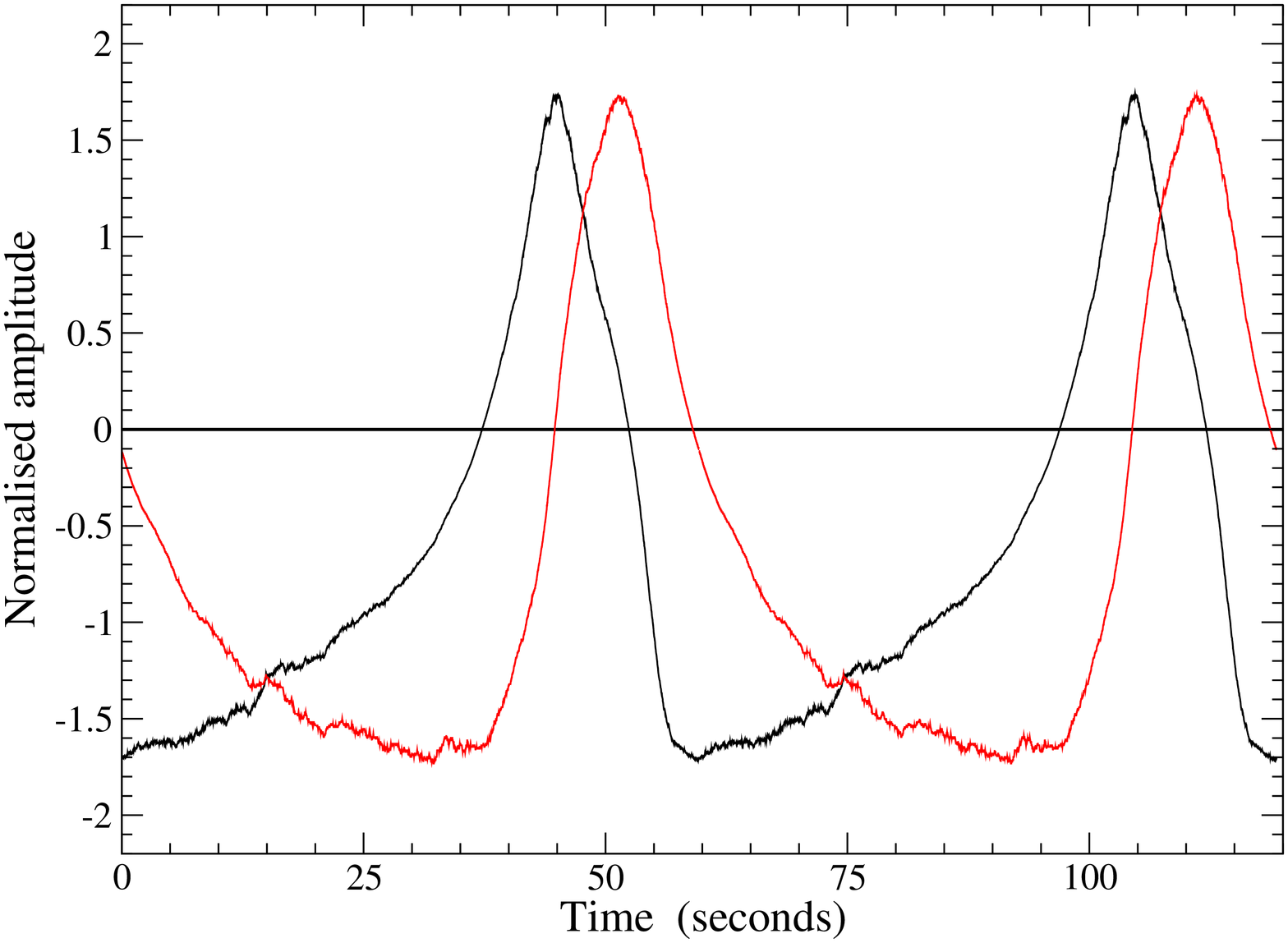}
\includegraphics[height=7.5cm, angle=-90]{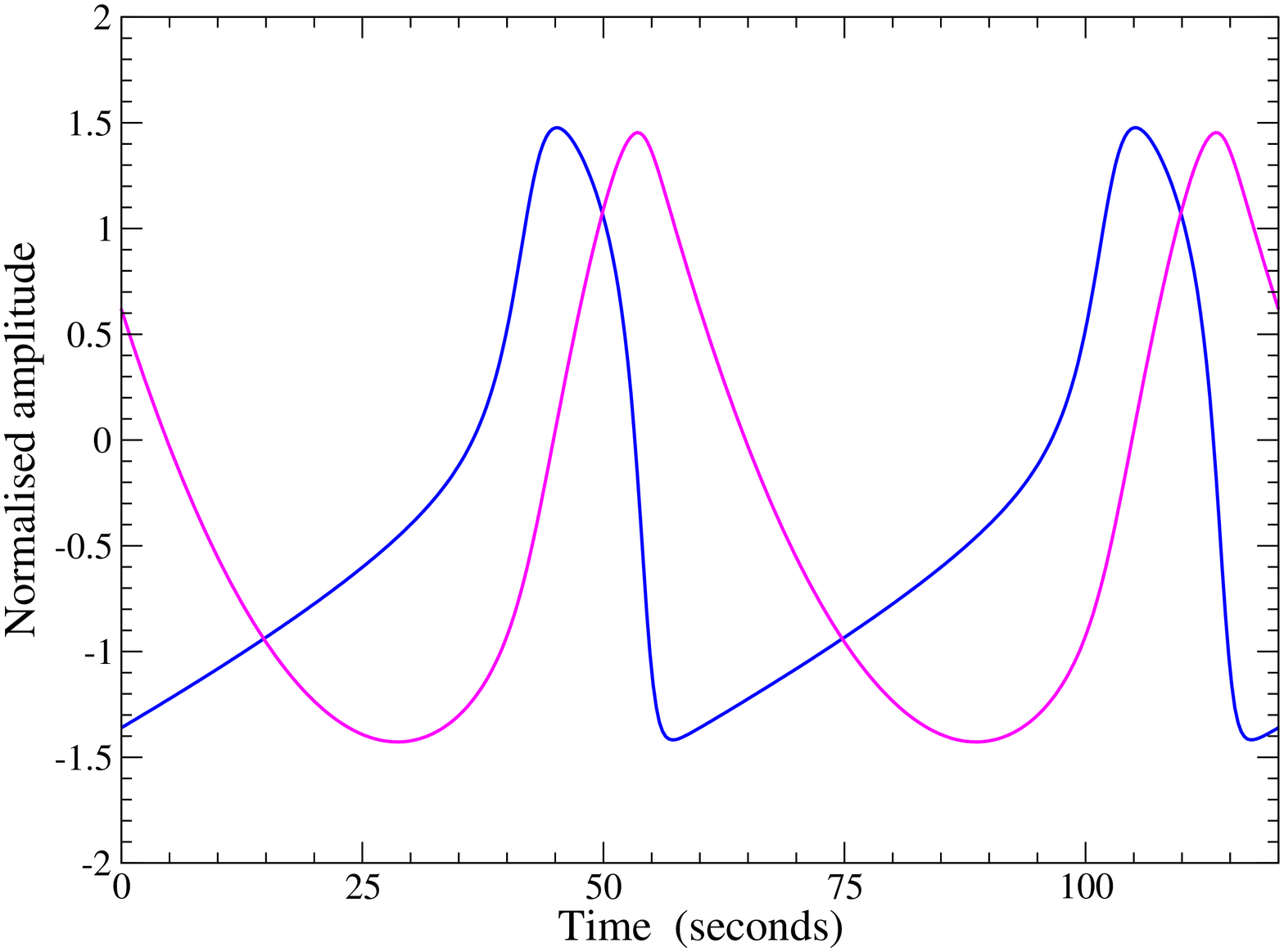}
\includegraphics[height=7.5cm, angle=-90]{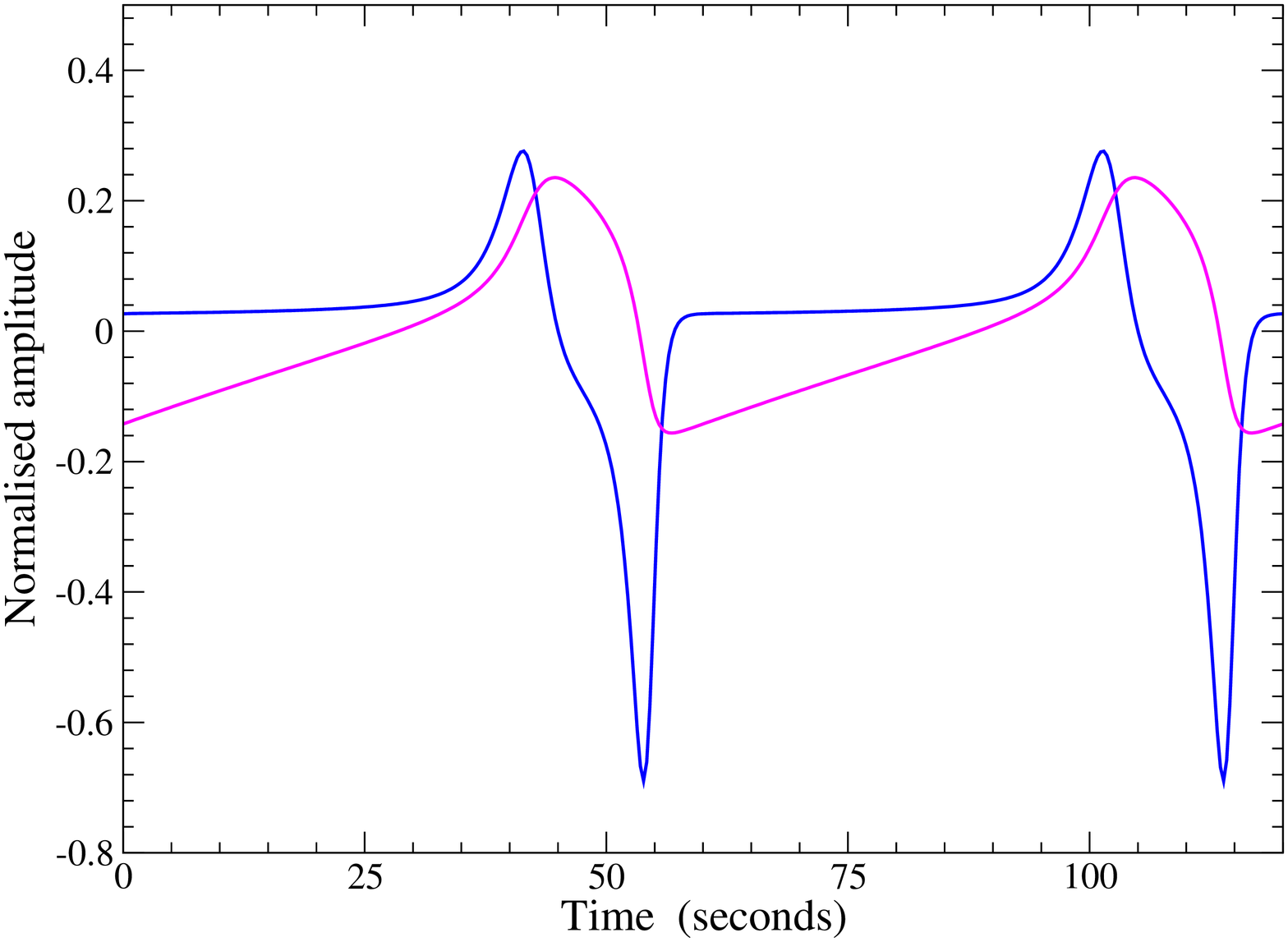}
\caption[]{
Top panel: time evolution of the mean pulse profile (black) and mean photon energy (red)
of the F7 series; 
Central panel: results obtained from the model FhN-A with the same parameters 
as in Fig.~\ref{fig5}, blue and magenta lines represent the time behaviour of the $x$ and $y$ 
variables, respectively using the same time scale of the upper panel,
Bottom panel: time derivatives of the $x$ and $y$ series shown in the central panel 
plotted using the same colors.
}
\label{fig6}
\end{figure}

A very important and unexpected finding of this solution is that the variable $y$
reproduces well the evolution of mean photon energy, or disk temperature 
\citep{Massa2013}.
The comparison between the mean burst and energy profiles with the $x$ and $y$ curves
is shown in the top and central panels of Fig.~\ref{fig6}.
Data and model curves were normalized, after subtraction of their central values, 
so to have comparable amplitudes. 
In the bottom panel of Fig.~\ref{fig6} we plotted the time derivatives of the two computed 
series to show more clearly how their changes occur on different time scales: 
the $y$ derivative (magenta curve) is always limited in a rather small interval, 
whereas that of $x$ remains small and positive during the $SLT$, dropping to 
a high negative value in the fast decaying phase of the \Puls and {\it FDT}.
One can also note a time delay between the two curves in the same direction of that 
in the observed data \citep{Massa2013}, although the latter resulted on the average 
smaller by about 25\%.

This finding makes stronger the heuristic meaning of the non-linear model and 
suggests a simple linear relation between the variable $x$ with the source 
photon luminosity (i.e. the rate of X-ray emitted photons) $L$ detected 
in the MECS energy range, or any other physical quantity proportional to 
it as the photon density in the emitting regions of the disk, and of $y$ with the mean 
energy of photons or the mean disk temperature, otherwise the main features of both data 
series would not be be so finely matched. 

\begin{eqnarray}
x(t) &=& C_L [L(t) - L_0] \nonumber \\
y(t) &=& C_E [E(t) - E_0]  
\end{eqnarray}

\noindent
where $L_0$ and $E_0$ are two constant values of the photon luminosity and mean photon 
energy, respectively, $C_L$ and $C_E$ are two dimensional coefficients useful 
for writing our equations in a non-dimensional form.
The correspondence between the formal variables $x$, $y$ and the physical observable 
quantities, is also apparent from the $x,y$ plot (hereafter phase space plot) that 
results similar, except for the scale factors, to the count-rate vs mean energy plot 
studied by \citet{Massa2013} \citep[see also Fig.14 in][]{Janiuk2005}.
This is clearly shown in Fig.~\ref{fig7}, where the loop of the stable cycle described by
Eqs.(6) is reported.

\section{Equilibrium points and stability}
The study of equilibrium points is important for understanding which are the conditions 
to develop a limit cycle behaviour \citep[see, for instance][]{Hale1991, Farkas1994,
Strogatz1994}.
Equilibrium points are obtained from the solutions of the system for 
$(dx/dt) = (dy/dt) = 0$, 
that for Eqs.(6) reads:

\begin{eqnarray}
 \gamma y &=& - \rho x^3 + \chi x - J    \nonumber \\
  x &=& y  
\end{eqnarray}
\noindent
that can be easily reduced to the cubic equation:

\begin{equation}
   \varphi (x) = x^3 + \frac{\gamma - \chi}{\rho}~ x + \frac{J}{\rho}  =  0
\end{equation}

\begin{figure}[tb]
\includegraphics[height=8.2cm,angle=-90,scale=1.0]{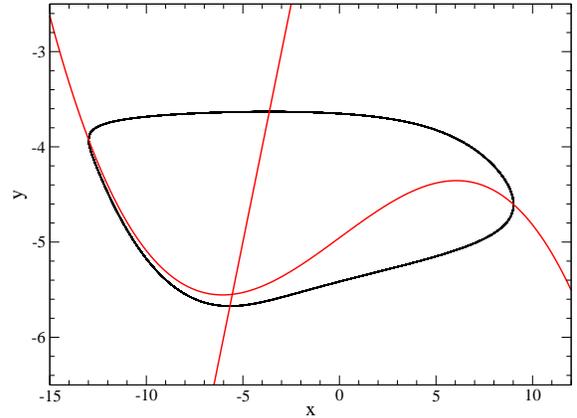}
\caption[]{
Phase space plot of the periodic solution of Eqs.(6) computed using the same
parameters' values of Fig.~\ref{fig5}. The loop is described in anti-clockwise direction
as observed in the count-rate vs mean energy plots discussed by \citet{Massa2013}.
Red lines are the {\it nullclines} for the model FhN-A.
}
\label{fig7}
\end{figure}

Light curves presented in Figs. 5 and 6 (FhN-A model) were obtained for 
$\chi / \gamma = 33/222  < 1$, and therefore the case of interest is the one 
with a unique (negative) equilibrium point, (see Appendix III), whose 
value for the considered parameters is $x_* = y_* = -$5.54891.
In the $x$,$y$ plane this equilibrium point is at the intersect between the $x = y$ 
line with the cubic curve $(dx/dt) =  - \rho x^3 + \chi x - J $ (these lines are 
named {\it nullclines}) which are plotted in red in Fig.~\ref{fig7}.
Note that for about all the $SLT$, in which the dynamical behaviour is ruled by the 
slow variable $y$, the trajectory remains very close to the $x$ nullcline, and only 
after it approaches the other nullcline near the equilibrium point, the fast variable 
turns to be dominant.
The trajectory rapidly moves to the right to reach the $x$ maximum on the other branch 
of the cubic line and then it decreases in a very short time towards the minimum on 
the former branch.

The results of the stability analysis of solutions in a neighborhood of this equilibrium 
point are given in the Appendix III.
The most interesting result is that around this equilibrium point a limit cycle can be 
established.

The equilibrium point corresponding to the parameters' values of the FhN-A model is quite
close to the stability boundary and a relatively small change of only one of these values 
can move the state outside this region, and as a consequence the limit cycle disappears.  
For example, keeping $\chi$ and $\rho$ fixed at the above values, the stability is
reached when $|x_*| > \sqrt{(\chi - 1)/3\rho} = 5.9628$, rather close to the equilibrium
value.
We will show in the next Section how the changes of the parameters affect the 
solution and whether these modifications can account for the observed variability 
of \grs.

\section{Effects of variations of parameters' values} 

One can use the numerical integration code to investigate how the burst shape and the 
recurrence time change with the parameters' values in Eqs.(6).
It is not simple to disentangle the role of individual parameters because  they 
are all combined together in determining the {\it nullclines} and the equilibrium point.
Therefore, in the following we will present the effects due to the change of only one 
of the parameters with the other three kept frozen to those of the model FhN-A.

\begin{figure*}[th!]
\includegraphics[height=8.2cm,angle=0,scale=1.0]{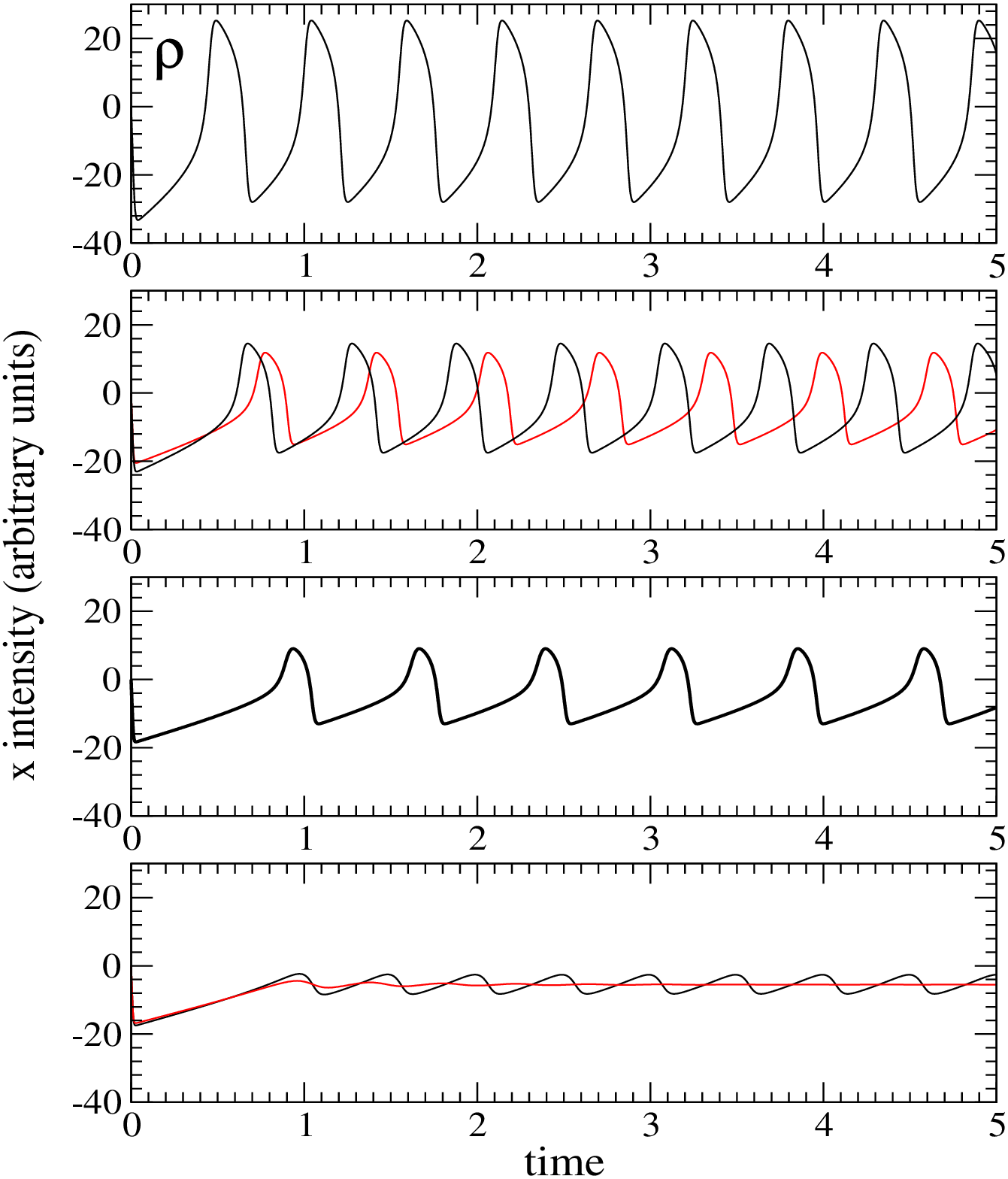}
\includegraphics[height=8.2cm,angle=0,scale=1.0]{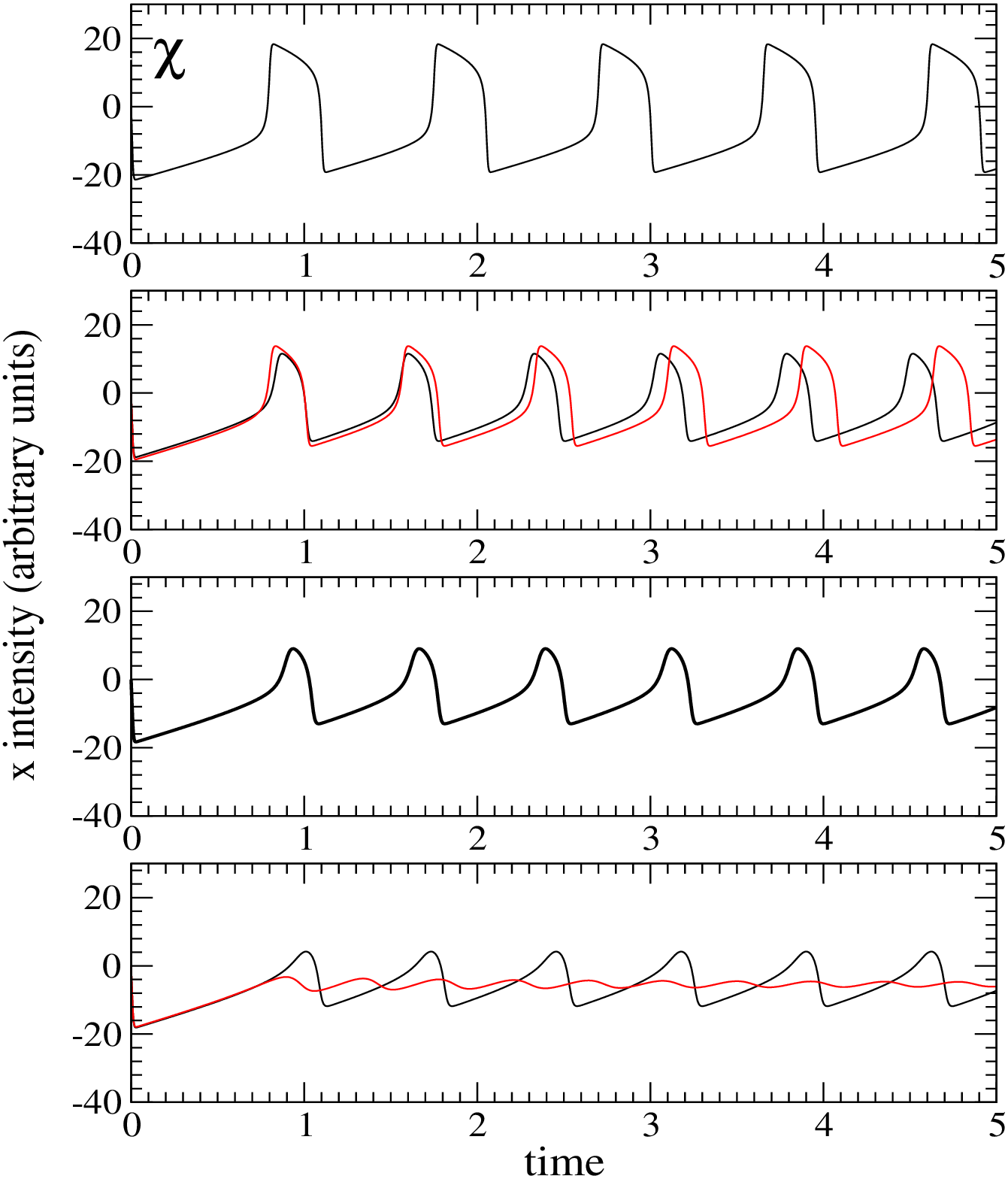} \\
\includegraphics[height=8.2cm,angle=0,scale=1.0]{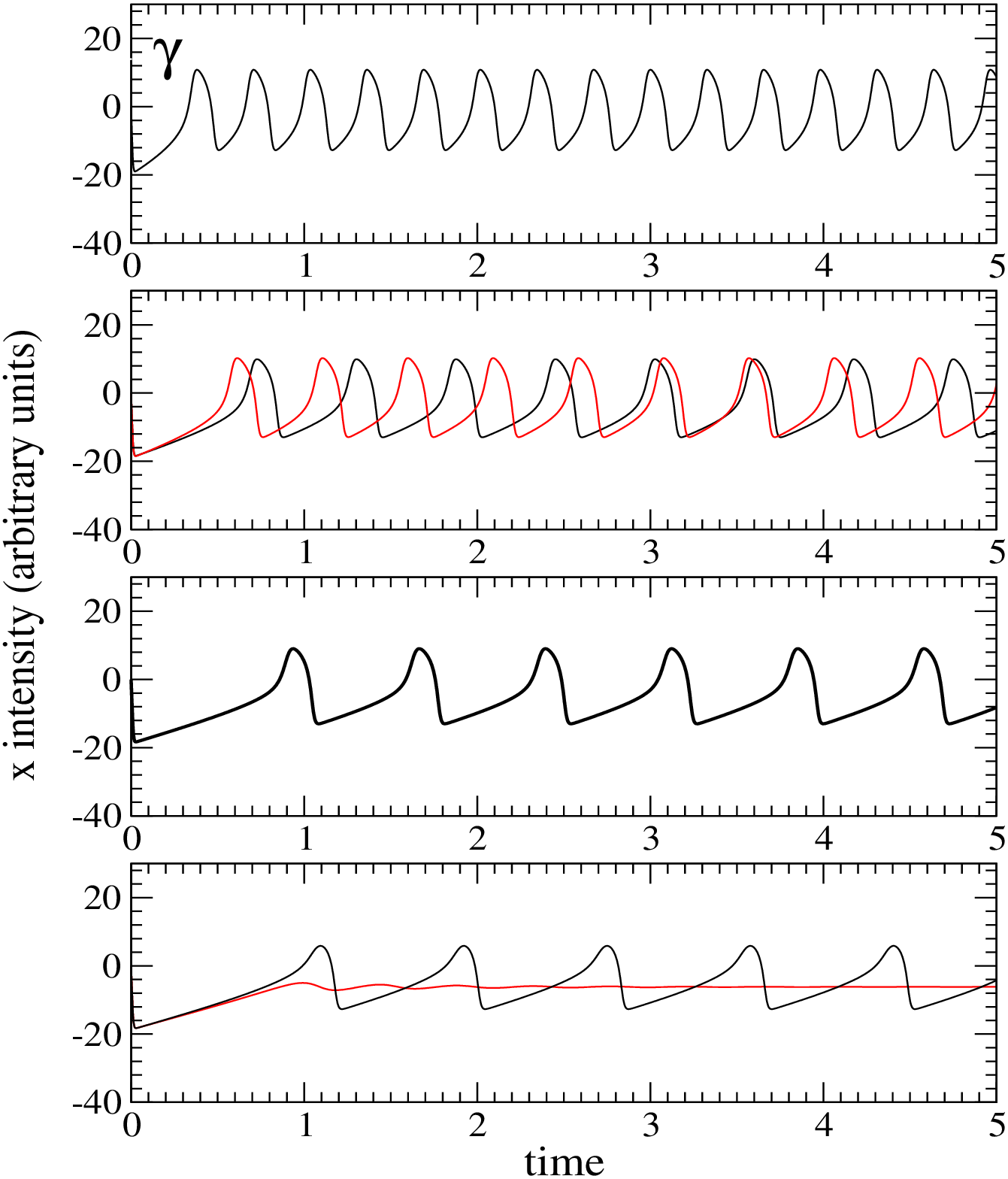}
\includegraphics[height=8.2cm,angle=0,scale=1.0]{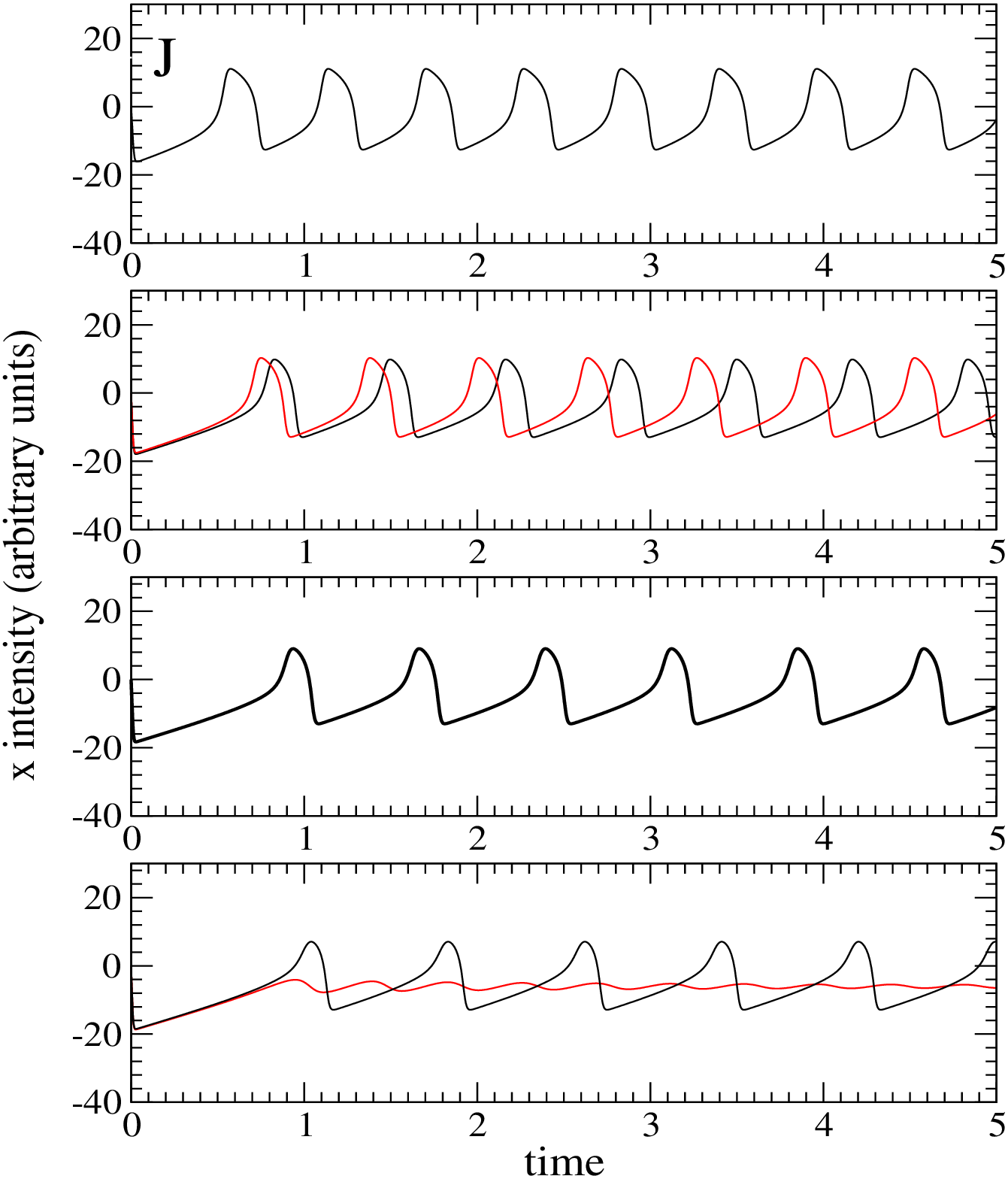}
\caption[]{
Variation of the burst period and amplitude for different values of the parameter $\rho$.
Top to bottom: 0.06, 0.16 (black) and 0.22 (red), 0.30, 0.34 (black) and 0.38 (red).
Variation of the burst period and amplitude for different values of the parameter $\chi$.
Top to bottom: 80.0, 60.0 (red) and 40.0 (black), 33.0, 30.0 (black) and 20.0 (red).
Variation of the burst period and amplitude for different values of the parameter $\gamma$.
Top to bottom: 500, 300 (red) and 260 (black), 222, 212 (black) and 200 (red).
Variation of the burst period and amplitude for different values of the parameter $J$.
Top to bottom: 600, 900 (red) and 1000 (black), 1100, 1150 (black) and 1200 (red).
}
\label{fig8}
\end{figure*}

The four panels in Fig.~\ref{fig8} summarize these results: they illustrate how a change of 
only one parameter can affect the light curve pattern. 
All the scales of the various panel are the same to make easy the comparison of the 
signals' profiles.
In each panel, the third curve from the top is the one of model FhN-A that we plotted 
to make easy the comparison.
A common characteristics is well evident: for all parameters there is a critical 
value for which the position of the equilibrium moves into a stability region
and the signal amplitudes decrease to a steady level.
These critical values are close to 0.34, 29.5, 210, and 1190 for $\rho$, $\chi$, 
$\gamma$ and $J$ respectively. 
In all these cases, they are remarkably close to the values found for the F7 series, 
indicating that during this time series, the system was close to switch-off the oscillations 
relaxing into a quasi-steady state.
Note also that variations of all the parameters can modify the burts recurrence time.

\subsection{$\rho$ changes}

Changes of the signal due to a variation of $\rho$ are shown in the let upper panel 
of Fig.~\ref{fig8}.
A rather small increase of $\rho$ from 0.30 to about 0.34 produces a switch off of 
the bursting and a stable level close to the mean value is maintained.
Stable burst patterns are obtained for decreasing $\rho$ down to quite small values 
around 0.05.
All the resulting curves show that a decrease of the \trec~ is associated with an 
increase of the amplitude. 
Modifications of the recurrence time produce also a shortening of the $SLT$ and an 
increase of the \Puls width with the consequence that the curve profile becomes more 
symmetric, with $r_s \approx r_p \approx 0.5$.

\subsection{$\chi$ changes}

The right upper panel of Fig.~\ref{fig8}  shows that a decrease  of $\chi$ from 33.0 to about 
30.0 is sufficient to reduce the burst amplitude by a factor of $\sim$ 2, and a 
further decrease will produce the almost disappearance of bursts with constant level remarkably 
close to the previous one.
An increase of $\chi$ produces an increase of $T_{rec}$ that appears mainly due to a 
broadening of the \Puls width instead of the $SLT$.

\subsection{$\gamma$ changes}

At variance with the two previous parameters, changes of $\gamma$ do not appear to affect 
significantly the burst amplitude but produce large variations of the burst recurrence 
time.
In the left lower panel of Fig.~\ref{fig8} one can see that the \Puls duration is practically 
unaffected by these changes, which affect mainly the length of the $SLT$.
This result agrees very well with the observed differences between A8b and F7 series.
Again we found that a relatively small decrease of $\chi$, from 222 to values lower than 
$\sim$ 210, would relax the bursting to a stable level, very close to those found above.

\begin{figure}[th!]
\includegraphics[height=8.2cm,angle=-90,scale=1.0]{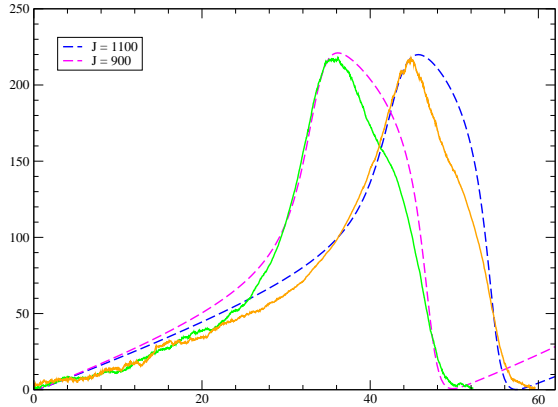}
\caption[]{
Comparison between the reconstructed pulse profiles of series F7 (orange) and A8b (green)
with the results of the FhN-A model for $J$=1100 (blue dashed line ) and 900 (magenta 
dashed line).
Time scaling is the same for the two computed series. 
}
\label{fig9}
\end{figure}

\subsection{$J$ changes}

Effects of $J$ changes are interesting because this is the only external parameter 
of the model.
Results are given in the right lower panel in Fig.~\ref{fig8}:
the \Puls width is practically unaffected by these changes whereas \trec~ appears to be 
much more variable, because of different durations of the $SLT$.
For low values of $J$ the curve profile tends to be symmetric, with a short $SLT$
and the corresponding values of $r_s$ and $r_p$ approach to 0.5.
The typical $\rho$ burst profile begins to be clearly recognizable for $J >$ 500, 
and approaches the observed one when $J$ is higher than 800.
The critical value of $J$ for the burst quenching is close to 1190. 

According to these results the main features of A8b series must correspond to a lower 
$J$.
We found that a $J$ value around 900 gives a pulse profile and \trec~ in a satisfactory
agreement with that of A8b series when the same scale factors of F7 data are used.
The shape of the \Pul, however, is slightly different from the observed one, being this
narrower than the one computed by the model.
A comparison between these results and the mean profiles is shown in Fig.~\ref{fig9}.

\begin{figure}[tb]
\includegraphics[height=8.2cm,angle=-90,scale=1.0]{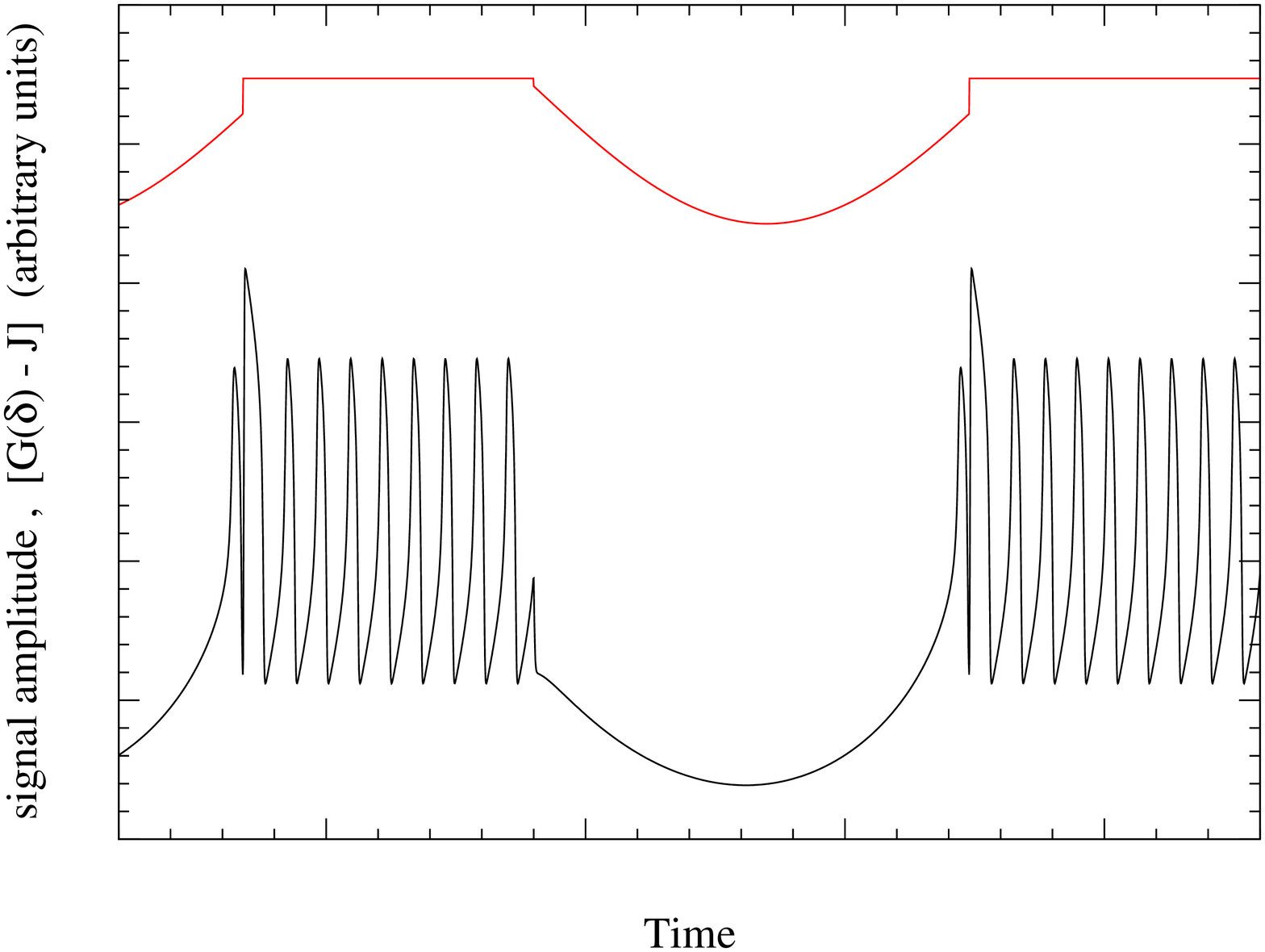}
\includegraphics[height=8.2cm,angle=-90,scale=1.0]{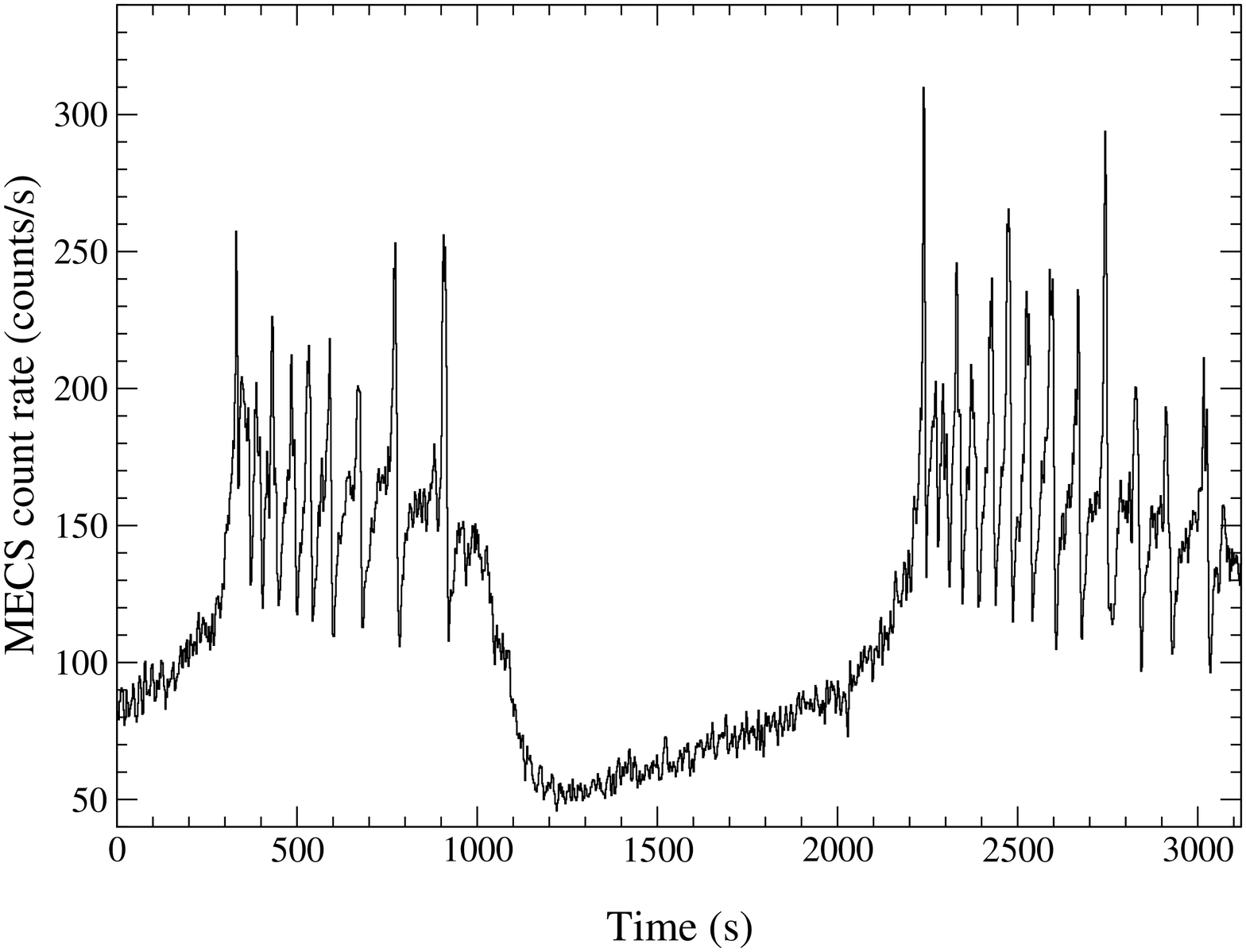}
\caption[]{
Upper panel: computed light curve (black line) when the $[G(\delta) - J]$ 
term changes as represented by the red line. Units are arbitrary.
Lower panel: \grss light curve in the $\nu$ class observed by MECS2 detector on 1996 
November 11. 
}
\label{fig10}
\end{figure}

\subsection{Slow changes of $J$}

It is very interesting to show how a nearly regular modulation of $J$ on time scales
longer than the typical recurrence time of bursts can produce different types of light 
curves resembling very much those of other \grss variability classes.
For instance, the upper panel of Fig.~\ref{fig10} shows what happens when $J$ has a slow modulation 
like that in the upper red curve.
Numerical calculations were performed by shifting both variables of $\delta = 24.0$ (see
Appendix II) to maintain always positive the forcing term $[G(\delta) - J]$.
The resulting light curve presents an alternance of bright states with a burst sequence
superposed and low brightness states, remarkably similar to those characterizing 
the $\nu$ class of \citep{Belloni2000}, as apparent from the light curve in the 
lower panel from a \sax~ observation of \grss performed on 1996 November 11. 

Note, in particular, that in addition to the main modulation other features agree with
observations, e.g. the first spikes reach higher count rates with respect to the following ones.
More realistic light curves are obtained including random fluctuations of $J$, 
as could be expected in a turbulent disk, that simulate either the observed noise and the 
small changes of \trec.

A spectral analysis of RXTE observations of \grss in the $\beta$ class was described by
\citet{Migliari2003} who reached the conclusion that the mass accretion rate is
variable with its highest value at the minimum brightness.
Results in Fig.~\ref{fig10} show that the $J$ must have a different behaviour with a
stable high value during the spikes' series followed by a slowly variable having a 
minimum practically simultaneous with the lowest brightness level.


\begin{figure}[tb]
\includegraphics[height=8.2cm,angle=-90,scale=1.0]{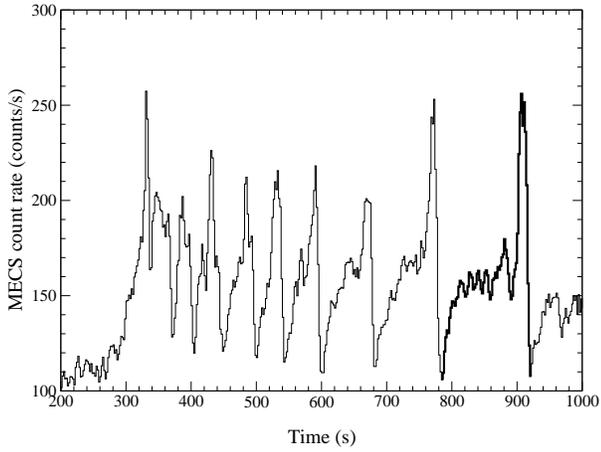}
\caption[]{Detail of the MECS2 light curve of \grss shown in the lower panel of Fig.~\ref{fig10}. 
The thick plotted burst is the same shown in Fig.~\ref{fig12}.
}
\label{fig11}
\end{figure}

\begin{figure}[tb]
\includegraphics[height=8.2cm,angle=-90,scale=1.0]{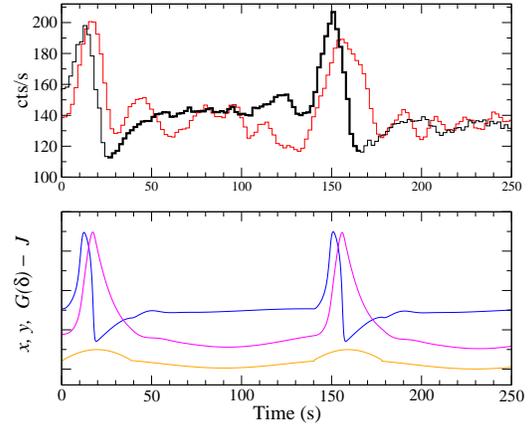}
\caption[]{
Upper panel: MECS2 light curve of a burst of the observation on 1996 November 11 (black),
red line is the corresponding mean photon energy scaled to the mean count rate level.
Bin width is 2 seconds, all data have been smoothed with a running average over 5 bins
to reduce the statistical noise.  
Thick line data are used to help the correspondence with data in Fig.~\ref{fig11}.
Lower panel: model light curve of the $x$ variable (blue line) obtained from the FhN-A 
model assuming a modulated $G(\delta) - J$ according to the orange curve.
The magenta line describes the variation of $y$ variable.
Time scale and amplitudes of these curves ware scaled to match the data in the upper
 panel.
}
\label{fig12}
\end{figure}

Variations of $J$ on time scales of the order a burst duration (typically $\sim$100 s)
can explain burst profiles as those of $\nu$ class (Fig.~\ref{fig4}). 
Fig.~\ref{fig11} shows a detail of the first burst series in Fig.~\ref{fig10}, where nine events are
clearly apparent.
Their duration increases more than a factor of 3, from $\sim$40 s to $\sim$ 140 s of 
the last one and the shape of some events has an initial well defined minimum followed 
by a fast increase turning in a rather flat shoulder which ends in a sharp spike
(Fig.~\ref{fig12} upper panel).
This pattern is obtained for the $x$ variable in FhN-A model (Fig.~\ref{fig12}  lower panel) 
assuming that $J$ has modulation as that shown by orange line.
In the same panel we reported also the solution for the $y$ (magenta curve) which 
corresponds well to the mean photon energy. 
The other three model parameters are unchanged demonstrating that the variation of 
only one quantity is able to reproduce all these behaviours.

An important point is that $J$ values producing the solution in Fig.~\ref{fig12}  vary in the 
range [1225,1390] with a mean value of 1330.
These values are higher than the one used to reproduce the $\rho$ class data and 
move the system into the locally stable region (see Appendix III, B region in Fig. A1).
Thus no limit cycle can exist and the photon luminosity changes must be originated from $J$ 
variations.
However, according to our model, it is not necessary that $J$ must have a modulation
amplitude comparable to that of the photon luminosity: a relatively small change can, in fact, 
produce much higher brightness changes because of the occurrence of non-linear effects.

\begin{figure}[tb]
\includegraphics[height=8.2cm,angle=-90,scale=1.0]{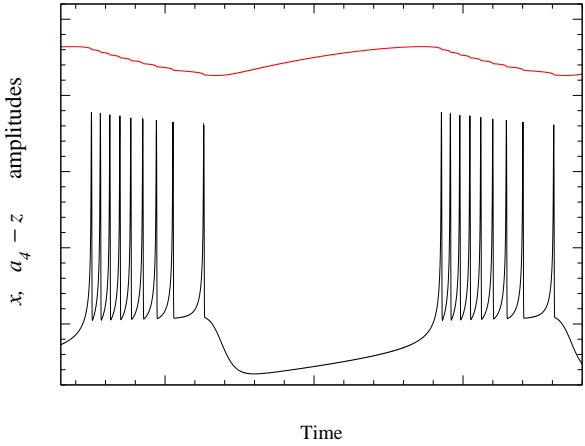}
\caption[]{Light curves ($x$ black line, $a_4 - z$ red line) computed by means of the HR 
model for reproducing the one observed by \sax~ and reported in the lower panel 
of Fig.~\ref{fig10}. 
Units are arbitrary.
}
\label{fig13}
\end{figure}

\section{Results from the Hindmarsh-Rose model}

In the FhN model time variations of a parameter must be taken as an input function and 
consequently the system is not autonomous.
This difficulty ca be solved by introducing a third variable, whose variations have a
on time scale much longer than those of $x$ and $y$, as in the case of the system 
proposed by \citet[][see Eq.(5) in Sec. 4]{Hindmarsh1984}.
We omit here the mathematical properties of the equations, and refer the reader to 
the above cited specific papers, and limit ourselves 
to present an interesting result obtained from numerical integration of Eqs. (5).
FhN-A results, of course, are easily obtained by HR model by taking $\varepsilon = 0$ 
and thus reducing the system only to two equations, but much more interesting are
the solutions when three equations are used.
In Eqs. (5) $z$ is the slow variable and plays a role similar to that of $J(t)$ 
in the previous Section, while $x$ and $y$ are the fast ones.
A possible light curve ($x$ variable) is shown in Fig.~\ref{fig10}, computed with 
the following values of the parameters: 
$\varepsilon = 0.001$, $a_1 = 1.0$, $a_2 = 3.0$, $a_4 = 1.9$, $a_5 = 5.0$,
$a_7 = -1.0$, $s = 4.0$ and $x_0 = -1.1$.
In the same figure it is also plotted the corresponding curve of the slow $a_4 - z$
variable (red line) which shows that it varies on the same scale of burst sequence, 
but not on that of individual spikes.
The similarity with the observed data in the lower panel of Fig.~\ref{fig10} is impressive:
in particular, non bursting phases exhibit very similar trends and also there is 
the very interesting property that the recurrence time of spikes during the bursting 
phase increases approaching the end of the series.
It is remarkable to mention that the same curve was early obtained by Hindmarsh \& Rose
(1984) for reproducing the behaviour of neurons.

It is interesting to mention how the solution changes with the value of $\varepsilon$:
when it decreases the length of burst sequence increases until it evolves in a long
and regular series of repeating spikes, while for increasing $\varepsilon$ the spike 
number decreases to unity giving again a regular series of spikes, but for values higher
than 0.0402 the solution changes and approaches rapidly a stable state.

\section{Discussion}

It is well known that the X-ray source \grs, the prototype of microquasars, 
exhibits a large number of variability classes changing from quasi-quiescent states 
to long and regular series of bursts, persisting for several days, and irregular 
variations on different time scales.
Many theoretical computations, generally based only on radial disk model equations, 
i.e. integrated over the thickness, and assuming the $\alpha$ prescription \citep{Shakura1976}
for the gas viscosity, have been performed in the past years 
since the early work by \citet{Taam1997}, with the goal of modelling these
puzzling light curves.

We followed a different approach and tried to reproduce some variability patterns 
of \grss by means of non-linear systems of differential equations, which have been 
extensively investigated in the literature in the contest of neuronal activity
to explain quiescent, bursting and spiking states.
Several systems of equations have been proposed, since the original work by  \citet{Hodgkin1952}.
We have shown that one of the simplest systems for neuronal model, developed by 
 \citet{Fitzhugh1961} from the Banhoffer-van der Pol oscillator and electronically realized 
by \citet{Nagumo1962}, consisting only of two differential equations with a single non-linear 
term, is able to reproduce some features of the $\rho$ variability class of \grs.  
We wrote the system in a form containing only four parameters and studied its equilibrium 
state and the stability conditions for which is possible to obtain a limit cycle.

In the following paragraphs we first summarize our results and then discuss a possible
interpretation in comparison with some proposed instability mechanisms.

\subsection{Summary of main results}

FitzHugh-Nagumo equations contain four parameters, three of which, $\rho$, $\chi$ 
and $\gamma$, relate the $x$ and $y$ time derivatives to the variables and are named
internal, while the only external parameter $J$ can play the same role of a forcing.
We adjusted parameters' values to obtain a solution for $x$ reproducing the gross 
profile of the $\rho$ class bursts and, unexpectedly, we found that the other variable 
$y$ exhibited a time modulation remarkably similar to that of the mean energy 
(or the temperature) of photons as found by \citet{Massa2013} (FhN-A model).
We underline that changes of $x$ and $y$ occur on two different time scales, fast and 
slow, respectively.
This result is obtained with a constant $J$ value and therefore it implies that this
bursting is intrinsic to the non-linear oscillations and do not require an external
modulation, as in the well known self-oscillation phenomenon (see for instance
the recent review paper by \citet{Jenkins2013}).
Moreover, the the delay of the emission at high energy with respect to the low energy 
one, as resulting from the plots in Fig.~\ref{fig6}, is a direct consequence of the 
physical mechanism responsible for the bursting and other mechanisms as photon scattering 
from a hot corona are not necessary, in agreement with the results of spectral analysis
\citep{Mineo2012}.
All these results agree remarkably well with the delay measurements reported by \citet{Massa2013}
and with the evolution of the disk and corona photon luminosity along the burst
obtained with the spectral analysis presented by \citet{Mineo2012}.
The FhN-A model, therefore, appears to have a relevant heuristic value and should not
be simply considered an ad-hoc mathematical description of the observed data.

It is also interesting to note that the non linearity of the FhN equations implies 
that the burst recurrence time is also depending upon the external parameter $J$.
The lower right panel in Fig.~\ref{fig8} shows that changes of \trecs are mainly due to a
modification of the {\it SLT} length.
From the curves in Fig.~\ref{fig9}, we see that the change of \trecs from A8b to F7 data 
series can be accounted for an increase of $J$ by about 20\%, comparable to that of
{\it BL} level in Fig.~\ref{fig1}.
This finding suggests that this parameter can be likely related to an external
mass-accretion rate ($\dot m$ at large radii) that is the main regulator of the disk 
photon luminosity.
Slow variations of this quantity and the consequent changes of $J(\dot m)$ can thus 
move the oscillator across the boundary between the stability or instability region, and 
{\it vice versa} (see Appendix III) with a consequent appearance or disappearance of 
a bursting limit cycle.   
Effects of $\dot m$ changes may also be responsible of some of the variability classes
as those classified by \citet{Belloni2000}. In Sect. 5.5 we found that a high and 
slowly modulated $J$ can produce solutions for $x$ with an alternance of spikes and
quasi quiescent states quite similar to those observed in the $\nu$ class light curves.

Under this respect, it appears more interesting the result obtained from the HR model,
which includes one more variable and a corresponding equation.
The bursting solution in Fig.~\ref{fig13} presents such a striking similarity with the data, 
such as the increasing recurrence time of bursts or the shape of the quiescent part
(see lower panel in Fig.~\ref{fig10}), that appears very unlikely that it is only a chance
result.
Our solution is essentially of the same type of those discussed by \citet{Shilnikov2008} 
and the mathematical properties are extensively presented in that paper.

We can conclude that the apparent complex behaviour of \grss seems to be mainly regulated 
by a single non-linear oscillator driven by a unique parameter, whose changes are
responsible at least for some of variability classes.
The non-linear oscillator model implies that there is an interplay between the two (or
three) variables and therefore some caveats must be considered in the interpretation of
spectral parameters' values derived by spectral fitting of the data, such as the widely used 
\textsf{diskbb} \citep{Mitsuda1984} or others in XSPEC. 
For instance, changes of the disk inner radius may alternatively be considered as a luminosity 
normalization rather that an evidence of a real change in the disk radial extension.

It is unclear  whether the models described in the present paper are the most appropriate 
ones for \grss or if other variables or terms must be introduced in the differential 
equations.
It is an open problem how a system of dynamical equations can be derived from the complex
mathematical apparatus of accretion disks and this requires new deep  theoretical 
investigations.
 
Useful indications in this direction, however, can be retrieved from some interesting
previously published works.

\subsection{Disk instabilities and the $\rho$ class limit cycle}

Studies of the development of instabilities in accretion disks started several decades 
ago \citep[e.g.][]{Lightman1974, Pringle1973, Shakura1976} and up to now originated an
 extensive literature.
\citet{Taam1984}, in particular, computed by means of numerical integration of
the non-linear disk equations to investigate thermal-viscous instabilities and obtained 
a few theoretical light curves having recurrent bursts consisting by a slow rising 
portion followed by a very narrow and high peak, more similar to those in the right 
panel of Fig.~\ref{fig4} or in Fig.~\ref{fig12}, rather than to those in Fig.~\ref{fig5}.
With the discovery of the $\rho$ class variability in \grs, \citet{Taam1997}
investigated the time and spectral properties of the bursts and proposed an 
interpretation based on the instability discussed in the previous paper.
However, these authors invoked for explaining the delayed hard emission a reflection 
in the frame of a disk-corona model.
Later \citet{Nayakshin2000} to reproduce some classes of light curves 
of \grss proposed a disk model with a rapidly variable viscosity and an upper  
limit to the energy fraction transferred to the disk emission.
\citet{Watarai2003} studied a model with a modified version of the $\alpha$-viscosity 
prescription law with respect to the standard one and introduced a power law dependence 
of viscosity stress tensor $\mathcal{T}_{r \varphi}$ on the ratio between the gas $p_{gas}$ 
and the total pressure $\beta = p_{gas} / (p_{gas} + p_{rad})$:
\begin{equation}
   \mathcal{T}_{r \varphi} = - \alpha_0 \beta^{\mu} \Pi
\end{equation}
where $\Pi$ is the height integrated pressure.

The resulting light curves exhibit the typical bursting behaviour and the variable 
recurrence time results from changes of the exponent $\mu$.
In these models, however, there is no delay between temperature and photon luminosity as 
observed in the $\rho$ class bursts \citep{Massa2013}.

A further investigation of the limit cycle light curves was performed by \citet{Merloni2006} 
who considered a magnetized disk in which magnetic turbulent stresses
inside the disk scale with the pressure as ($0 < \mu < 2$): 
\begin{equation}
   \mathcal{T}_{r \varphi} = - \alpha_0 p_{tot}^{1 - \mu/2} p_{gas}^{\mu/2}
\end{equation}

Their calculated light curves for an accreting black hole of 10 $M_{\odot}$
and $\mu = 0.1$ have a recurrence time of bursts increasing with the mass accretion 
rate, in a qualitative agreement with our results.
Unfortunately, the simultaneous temperature evolution is not given and a full
comparison with the data and the results of the FhN-A model is not possible.
However, the approach of an energy transfer instability in a magnetized disk appears
one of the most promising for the understanding of physical processes regulating
the instability.

A model for the $\rho$ class has been recently proposed by \citet{Neilsen2012, Neilsen2011} 
who interpreted the bursting as a consequence of a similar modulation in the
mass accretion rate.
In these works, authors used RXTE/PCA and Chandra/HETGS data to characterize the spectral
 modifications according to the burst phase. A model of soft thermal disk component and 
hard Comptonized emission shows that changes in the parameters are smooth for most of the 
burst phases, but a significant spectral modification is observed during the pulse. At this 
phase it is observed a strong decrease of the inner disk radius (up to $\sim$ 2 R$_{g}$, that 
would require a spinning black-hole with an adimensional angular momentum per unit mass
$a \simeq 0.9$ if this minimal radius is identified with that of the last stable orbit), 
and a strong increase in the coronal electron opacity (where most of the disk emission is 
Thomson scattered). 
Some authors \citep[e.g.][]{Artemova1996} have therefore argued that during the pulse 
a recurrent set of steps are operating: when at small disk radii, local Eddington limit is 
reached, a radiation pressure instability develops and, consequently, part of the inner disk 
is vaporized into an optically thick, completely ionized, cloud. This cloud emits also part 
of its energy as thermal bremsstrahlung emission (hard pulse). Modification of this primary 
radiation causes also changes at large outer disk radii, as it strongly modified the structure 
of any (thermally/radiation driven) disk wind. Because the fraction of the mass outflow that 
is dispersed into a wind (which if channelled by the disk magnetic field lines may become a jet) 
modifies the fraction of the accretion mass rate, this leads to generation of mass density 
waves that propagate toward the accretor, giving rise to the burst typical recurrent pattern. 
Such scenario appears to be consistent with  a combination of spectroscopic detailed line diagnostic and 
broadband coverage of the spectral shape. However, our approach shows that the limit-cycle 
behaviour can also be understood independently from any fitted spectral decomposition. If 
the strong relation of the global mass accretion rate with $J$ holds, the link between the 
outflow rate with the inner mass accretion rate could not be  strictly required. 

We have shown that  the outer mass accretion rate is probably responsible for passages 
between one state to an  another, but within the $\rho$-class state, $J$ is constant, 
and the oscillating mechanism is still at work. 
This does not mean that changes at small radii should not impact the wind 
physical properties, but only that the claimed strong feed-back of the two zones may not be 
necessary for the development of the limit-cycle instability. We note that a detailed modelling 
of the expected changes in the energy spectra would require a translation of the differential 
equations into the proper physical radiation components, which would lead beyond the aims of 
the present work.  

\acknowledgments
The authors are grateful to Marco Salvati and Andrea Tramacere for very 
useful discussion and comments.
The CNR Institutes and the \sax~ Science Data Center
are financially supported by the Italian Space Agency (ASI) in the
framework of the \sax~ mission.

\bibliographystyle{spr-mp-nameyear-cnd}
\bibliography{grs1915}

\appendix{

\section*{Appendix I - Derivation of FhN equations}

Consider the two equations of the general system given of Eq. (4) written using the
variables $\widetilde{x}(\widetilde{t})$ and $\widetilde{y}(\widetilde{t})$:
\begin{eqnarray}
\frac{d\widetilde{x}}{d\widetilde{t}} &=&  \frac{1}{A} [ -a_1 \widetilde{x}^3 + a_3 \widetilde{x} - b_1 \widetilde{y} ] \nonumber \\
\frac{d\widetilde{y}}{d\widetilde{t}} &=& a_6 \widetilde{x} - b_2 \widetilde{y} + a_7  \nonumber
\end{eqnarray} 

\noindent
and transform it by means of the following change of variables
$x = (a_6/b_2) \widetilde{x}$,~ $y = \widetilde{y} - (a_7/b_2)$, and $t = b_2 \widetilde{t}$.

After some simple calculations one obtains:
\begin{eqnarray}
\frac{dx}{dt}&=& - \frac{a_1 b_2}{A a_6^2} x^3 + \frac{a_3}{A b_2} x -\frac{a_6 b_1}{A b_2^2} y  - \frac{a_6 a_7 b_1}{A b_2^3} \nonumber \\
\frac{dy}{dt} &=& x - y  \nonumber
\end{eqnarray}

\noindent
and introducing the following coefficients 
\begin{eqnarray}
\rho &=&  \frac{a_1 b_2}{A a_6^2}    \nonumber \\
\chi &=&  \frac{a_3}{A b_2}          \nonumber \\
\gamma &=&  \frac{a_6 b_1}{A b_2^2}  \nonumber \\
J  &=&  \frac{a_6 a_7 b_1}{A b_2^3} ~~~~~~~~~ \nonumber 
\end{eqnarray}

\noindent
it becomes the same system of Eq. (6).

\section*{Appendix II - Invariant translated equations}

Eqs.(6) are written in a particular simple form having a rather small number of
parameters (three internal parameters and an external one).
In this approach, however, the physical meaning of the parameter would not result
always clear: for instance the external term $J$, that we related with the mass 
accretion rate, gives a negative contribution to the $x$ derivative in the former
of Eqs.(6), and this can appear to be unphysical.

One has to consider that the shape of Eqs.(6) solutions is invariant under 
a translation along the fist degree nullcline. Shifting both $x$ and $y$ of
the same quantity $\delta$, we can define the new variables $\xi= x + \delta$ 
and $\eta = y + \delta$, whose time derivatives are obviously equal to those
of $x$ and $y$, respectively.
Eqs.(6) transform into the more general ones: 
$$
\frac{d\xi}{dt} = - \rho \xi^3 + \chi \xi - \gamma \eta + 3 \rho \delta (\xi^2 -\delta \xi)  
  + [G(\delta) - J] ~~~~~~~~~ \nonumber \\
$$
$$
\noindent
\frac{d\eta}{dt} = \xi - \eta ~~~~~~~~~~~~~~~~~~~~~~~~~~~~~~~~~~~~~~~~~~~~~~~~~~~~~~~~~~~~~ 
$$

\noindent
where
$$
\noindent
G(\delta) = 3 \rho \delta^3 -  \chi \delta + \gamma \delta ~~~~.~~~~~~~~~~~~~~~~~~~~~~~~~~~~~~~~~~~~~~~~~~~~~~~~~~~~~~~ 
$$

It is rather simple to verify that the term in square brackets for a $\delta$ 
value large enough can obtain a positive value, as expected. 
Solutions of this system differ from those of the original Eqs.(6) only for an 
additive constant, thus the quantities  $\xi - \delta$ and $\eta - \delta$ are
invariant.
It is clear that only the case $\delta = 0$, without the quadratic term, is more
suitable for the study of nullcline properties, and for this reason we considered 
it in our work.

\section*{Appendix III - Stability and limit cycle conditions}

Here we present a synthetic description of main results concerning the stability 
of equilibrium points and the conditions to have a limit cycle within a limited 
region surrounding them.
A complete mathematical description will be given in a separate paper (Ardito et 
al., in prepatation).
The equilibrium points are the solutions of the cubic Eq. (9) 
$$\noindent
   \varphi (x) = x^3 + \frac{\gamma - \chi}{\rho}~ x + \frac{J}{\rho}  =  0~~~~~~~~~~~~~~~~~~~~~~~~~~~~~~~~~~~~~~
$$

Three cases must be considered:
\noindent
$i)$ $ \chi < \gamma + 3 (\rho J^2 / 4)^{1/3}$,
there is only a unique real solution $x_* < 0$ and therefore only one 
equilibrium point ($x_*, y_*$) exists;

\noindent
$ii)$ $\chi = \gamma + 3  (\rho J^2 / 4)^{1/3}$,
there are a negative solution $x_* < - \sqrt{(\chi - \gamma)/(3 \rho)}$ 
and a double positive solution $x_0 = \sqrt{(\chi - \gamma)/(3 \rho)}$, 
the system has two equilibrium points;

\noindent
$iii)$ $\chi > \gamma + 3  (\rho J^2 / 4)^{1/3}$,
there are three real solutions $x_*, x_1, x_2$, with 
$x_* < - \sqrt{(\chi - \gamma)/(3 \rho)} < x_1 < \sqrt{(\chi - \gamma)/(3 \rho)} < x_2$,
and therefore the system has three equilibrium points.

Considering that $x_*$ is the unique negative solution of the equation 
$\varphi(x) = 0$, in the following we will use $x_*$ as parameter instead of $J$.

For the study of the stability it is more convenient to define the two 
new variables:
\begin{eqnarray}
  u &=& x - x_*      \nonumber \\
  v &=& (y - y_*) - (x - x_*)/\gamma   \nonumber 
\end{eqnarray}
\noindent
and the system of differential equation becomes:
\begin{eqnarray}
  du/dt &=& \gamma [f(u) -v]      \nonumber \\
  dv/dt &=& g(u)           \nonumber       
\end{eqnarray}
\noindent
where \\
\noindent
$f(u) = - (\rho/\gamma) u [u^2 + 3x_* u + 3 x_*^2 + (1/\rho) -(\chi/\gamma)] $ \\
\noindent
and \\
\noindent
$g(u) =  (\rho/\gamma) u [u^2 + 3x_* u + 3 x_*^2 + (\gamma/\rho) -(\chi/\gamma)] $ ~~, \\
\noindent
with $\chi,~\rho, ~\gamma > 0$, $x_* < 0$, and $ (\chi / \rho) < (\gamma / \rho) + x_*^2 $.  

The previous conditions are now written as:

\noindent
$i'$) for $ 0 < \chi / \rho < \gamma / \rho + (3/4) x_*^2$ there is only an 
equilibrium point at the origin $O$ in the $(u, v)$ space;

\noindent
$ii'$) for $ \chi / \rho = \gamma / \rho + (3/4) x_*^2$ there are two 
equilibrium points, one at the origin $O$ and the other at 
$E_0 =$($-(3/2) x_* ; -(3/2) (1 - 1 / \gamma) x_*$) in the $(u, v)$ space;

\noindent
$iii'$) $ \gamma / \rho + (3/4) x_*^2 < \chi / \rho < \gamma / \rho + x_*^2$ 
there are three equilibrium points $O$, $E_1 = (u_1, v_1)$, $E_2 = (u_2, v_2)$, 
where
$$
u_{1,2} = -(3/2) x_* \mp \sqrt{ (\chi / \rho) - (\gamma / \rho) - (3/4) x_*^2}
 ~~~~~~~~~~~~~~~~~~~~~~
$$
$$
v_{1,2} = (1 - 1 / \gamma) u_{1,2}
 ~~~~~~~~~~~~~~~~~~~~~~~~~~~~~~~~~~~~~~~~~~~~~~~~~~~~~~~~~~~
$$

In the case $\gamma > 1$, that is the interesting one for us, 
from the linear analysis \citep{Hale1991} one obtains the following
results:

\noindent
$i'$) 

for $\chi / \rho < 1 / \rho + 3 x_*^2$, $O$ is locally asymptotically stable;

for $\chi / \rho = 1 / \rho + 3 x_*^2$, $O$ is linearly stable;

for $\chi / \rho > 1 / \rho + 3 x_*^2$, $O$ is unstable;

\noindent
$ii'$) 
$E_0$ is always unstable, whereas for $O$ we obtain the following conditions:

for $ 0 < 1 / \gamma < 1 -(9\rho /4\gamma) x_*^2$, $O$ is unstable;

for $ 0 < 1 / \gamma = 1 -(9\rho /4\gamma) x_*^2$, $O$ is linearly stable;

for $ 1 / \gamma > 1 -(9\rho /4\gamma) x_*^2$, $O$ is locally asymptotically 

stable;

\noindent
$iii'$) 
$E_1$ is always unstable, $E_2$ is locally asymptotically stable, whereas for $O$ we 
obtain the following:

for $\chi / \rho > {\rm max}( \rho / \gamma + (3/4) x_*^2, (1 / \rho) + 3 x_*^2 )$, $O$ is 
unstable;

for $\chi / \rho = (1 / \rho) + 3 x_*^2$, $O$ is linearly stable;

for $ \gamma / \rho + (3/4) x_*^2 < \chi / \rho < (1 / \rho) + 3 x_*^2$, $O$ is locally 
asymptotically stable.

It is useful to consider the 
space of the two parameters $\chi / \rho$ and $x_*^2$, represented in the two panels 
of Fig. A1.
Here the black, green and red lines are given by the respective equations:
$$
 (\chi / \rho) = (\gamma / \rho) +  x_*^2 
~~~~~~~~~~~~~~~~~~~~~~~~~~~~~~~~~~~~~~~~~~~~~~~~~~~~~~~(A1)
$$
\noindent
$$
 (\chi / \rho) = (\gamma / \rho) + \frac{3}{4} x_*^2 
~~~~~~~~~~~~~~~~~~~~~~~~~~~~~~~~~~~~~~~~~~~~~~~~~~~~~~~(A2)
$$
\noindent
and
$$
 (\chi / \rho) = ( 1 / \rho) + 3 x_*^2 
~~~~~~~~~~~~~~~~~~~~~~~~~~~~~~~~~~~~~~~~~~~~~~~~~~~~~~~(A3)
$$

\noindent
where the first equality is the upper limit of acceptable values of $\chi / \rho$.

\begin{figure}[tb]
\includegraphics[height=8.2cm,angle=-90,scale=1.0]{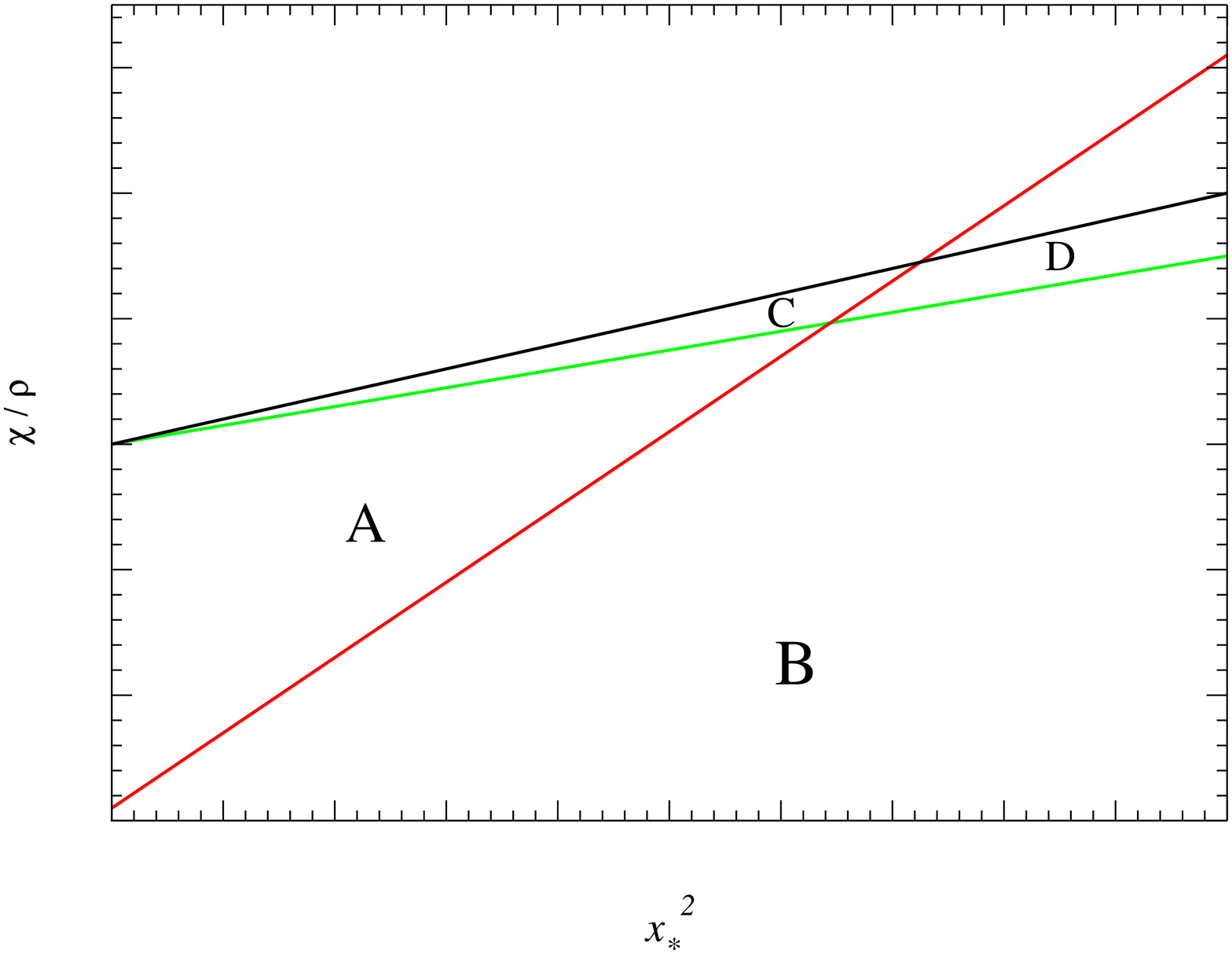}
\includegraphics[height=8.2cm,angle=-90,scale=1.0]{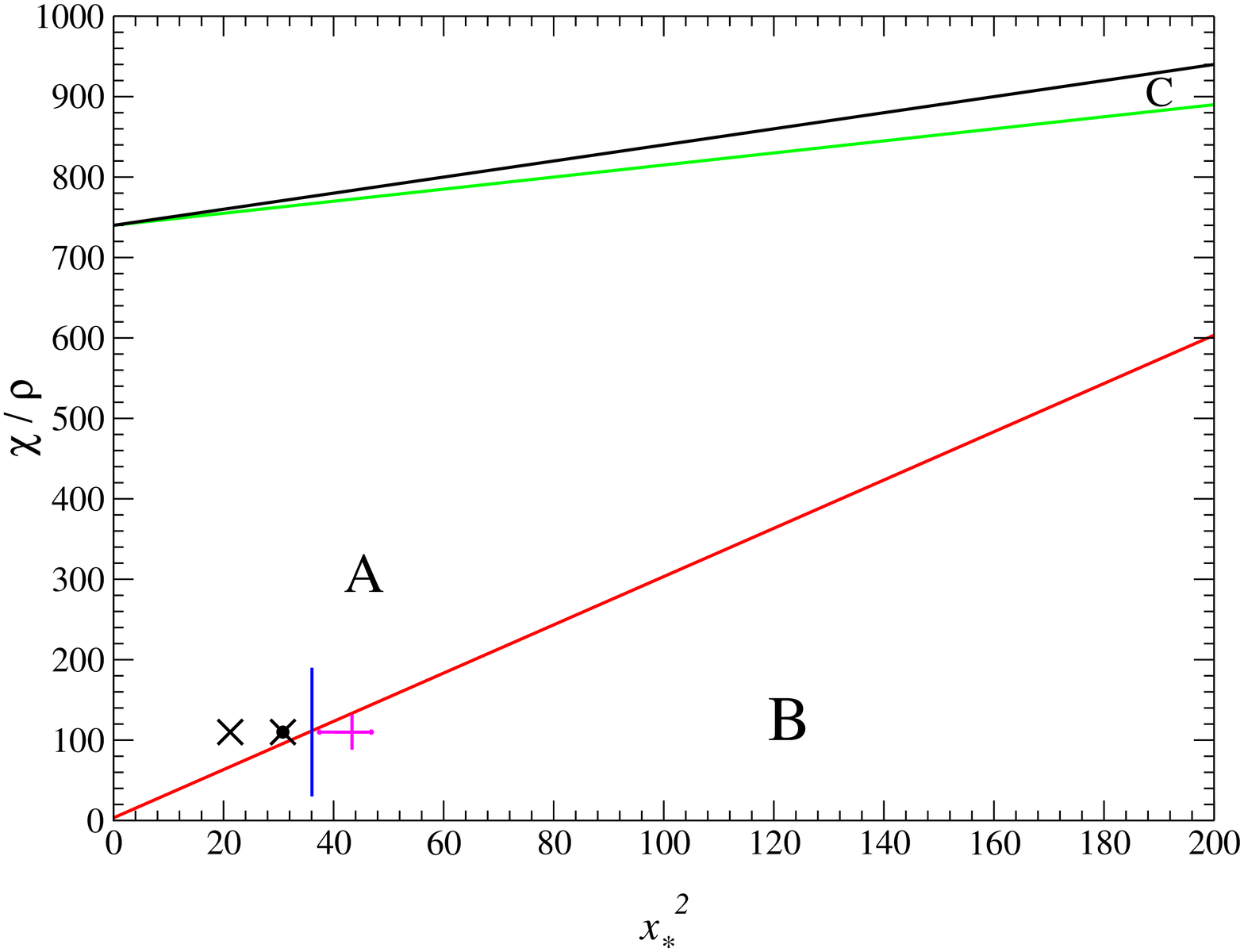}
\caption[]{Stability plot in the plane $(\chi / \rho)$ vs $x_*^2$, for
the FhN-A model. 
Upper panel:
black, green and red lines correspond to equations (A1), (A2) and (A3), respectively,
which define the four regions A, B, C and D in which solutions have different
stability, as explained in the text. Units are arbitrary.

Lower panel:
the same plot for the model FhN-A describing the $\rho$ class bursts:
the cross with a circle marks the point corresponding to the numerical solution 
for the F7 series ($J$ = 1100, see Sect. 3.1) and the simple cross the 
solution for the A8b series ($J$ = 900); the blue line corresponds to the equilibrium
point for $J =1190$, for which the transition to stability occurs; the magenta
line is the interval where is obtained the solution plotted in Fig.~\ref{fig13}. 
}
\end{figure}

We obtain thus four regions, marked by the letters A, B, C, and D where the system 
stability is:
\begin{itemize}
\item
region A: there is only one unstable equilibrium point;
 
\item
region B: there is only one equilibrium point which has asymptotic local stability; 

\item
region C: there are three equilibrium points, two of which are unstable and the third
has an asymptotic local stability;

\item
region D: there are three equilibrium points, one is unstable and the other two have
asymptotic local stability.
\end{itemize}

The lower panel in Fig. A1 shows the same plane for the numerical solutions we obtained
for the A8b and F7 series. 
The corresponding points in this plane are indicated by a cross: they lie in region A, 
and the latter is close to the boundary line with stable region B. 
A small change in the value of $J$ can move the equilibrium point across the red line 
into the stable region with a disappearance of the bursting.
The vertical blue line marks the value of $x_*^2$ for $J$=1190, for which we have the 
transition to a stable solution.

In the third case of $i'$, that corresponds to the region A in Fig. A1, we have
$u ~g(u) > 0$, $\forall u \ne 0$ and 
$ f(u) = -(\rho / \gamma) u (u - \widetilde{u}_1)(u-\widetilde{u}_2)$, with
$\widetilde{u}_{1,2} = -(3/2) x_* \mp \sqrt{ (\chi / \rho) - (1 / \rho) 
- (3/4) x_*^2}$.

Taking $\forall \theta \in [0, \sqrt{ \rho ((3/4) x_*^2 + (\gamma / \rho) 
- (\chi / \rho)}~)~]$,
we can consider the family of Lyapunov functions \citep{Farkas1994}:
$$
V_{\theta}(u,v) = \frac{1}{\gamma} \int_0^u g(s) d s + \frac{\theta}{\gamma} 
u v + \frac{v^2}{2}  
~~~~~~~~~~~~~~~~~~~~~~~~~~~.
$$  

Let us observe that $\forall c > 0$ the equation $ V_{\theta}(u,v) = c $ defines a simple 
closed curve that is the boundary of the domain 
$$ \mathcal{D}(\theta, c) = \{(u, v) \in  \mathbf{R}^2 | V_{\theta}(u,v) < c \}
~~~~~~~~~~~~~~~~~~~~~~~~~~~~,$$
which includes the equilibrium point $O$.

It is possible to demonstrate that for $\theta = 0$ and for every initial point 
$(\bar{u}, \bar{v}) \ne (0, 0)$ it exists $\bar{t} > 0$ such that for $t > \bar{t}$
we have that $(u(t,\bar{u}), v(t,\bar{v}))$ 
lies outside the domain $\mathcal{D}(0, c_0)$
where
$$ c_0 = \frac{1}{\gamma} \int_0^{\widetilde{u}_1} g(s) d s 
~~~~~~~~~~~~~~~~~~~~~~~~~~~~~~~~~~~~~~~~~~~~~~.$$

Moreover, because the derivative of $V_{\theta}(u,v)$ tends to $-\infty$ for 
$||(u,v)|| \to \infty$, one has that at least one $c(\theta) > 0$ exists such 
that, for $\forall c > c(\theta)$ it exists $\widetilde{t} > 0$ such that for $t > \widetilde{t}$
we have that 
$(u(t,\bar{u}), v(t,\bar{v})) \in \mathcal{D}(\theta, c)$.
 
Finally, for 
$$\bar{c} > c_0 + (\theta / \gamma) 
\big( {\rm max}\big\{uv~ |~ (u,v) \in \overline{D}(0,c_0)\big\} \big)~~~~~~~~~~~~~~~~~~~~~~~,$$
it follows that
$$ \mathcal{D}(0, c_0) \subset \mathcal{D}(\theta,\bar{c})
~~~~~~~~~~~~~~~~~~~~~~~~~~~~~~~~~~~~~~~~~~~~~~~~~~~~~~~~~$$
that implies the existence of at least one periodic solution.

}
         
\end{document}